\documentclass[preprint,aps,prc,showpacs,nofootinbib,floatfix]{revtex4}

\usepackage{graphicx}
\usepackage{bm}

\newcommand{\clebsch}[6]{\left\langle  #1\, #2 ; #3\, #4 \vert #5\, #6 \right\rangle^{2}  }
\newcommand{\Pythia}{\textsc{Pythia}}
\newcommand{\Fritiof}{\textsc{Fritiof}}

\begin{document}

\title{Kaon and hyperon production in antiproton-induced reactions on nuclei}

\author{A.B. Larionov$^{1,2}$\footnote{Corresponding author.\\ 
        E-mail address: Alexei.Larionov@theo.physik.uni-giessen.de}, 
        T. Gaitanos$^1$, and U. Mosel$^1$}

\affiliation{$^1$Institut f\"ur Theoretische Physik, Universit\"at Giessen,
             D-35392 Giessen, Germany\\
             $^2$National Research Center ``Kurchatov Institute'', 
             123182 Moscow, Russia}

\date{\today}

\begin{abstract}
We study the strangeness production in antiproton-nucleus collisions 
at the beam momenta from 200 MeV/c to 15 GeV/c and in $\bar p$ annihilation
at rest within the Giessen 
Botzmann-Uehling-Uhlenbeck (GiBUU) transport model. 
The GiBUU model contains a very detailed description
of underlying antinucleon-nucleon cross sections, in-particular, of the strangeness 
production channels.
We compare our calculations with experimental data on $\Lambda$, $K^0_S$ and $\bar\Lambda$ 
production in $\bar p$A collisions and with earlier intranuclear
cascade calculations. The contributions of various partial channels
to the hyperon production are reported and systematic differences with the 
experiment are discussed. The possible formation of bound  $\Lambda$- and 
$\Lambda\Lambda$-nucleus systems is also considered. Finally, results on
the $\Xi$ hyperon production are presented in view of forthcoming 
experiments with antiproton beams at FAIR.  
\end{abstract} 


\maketitle

\section{Introduction}
\label{Intro}

The production of strange particles in antiproton annihilation on nuclei has been a challenge to
theory since about two decades when the $\Lambda$ and $K^0_S$ yields and spectra
were measured in the series of bubble chamber experiments 
\cite{Condo:1984ns,Balestra:1987vy,Miyano:1988mq}. The most intriguing experimental fact is 
that the yield ratio $\Lambda/K^0_S$ is very large $(> 2)$ not only for heavy target nuclei
($\bar p + ^{181}$Ta at 4 GeV/c \cite{Miyano:1988mq}), but even for light ones 
($\bar p + ^{20}$Ne at 608 MeV/c \cite{Balestra:1987vy}). The direct mechanism $\bar p p \to \bar\Lambda \Lambda$
can only explain less than 20\% of the measured $\Lambda$ production cross section 
for $\bar p + ^{181}$Ta at 4 GeV/c \cite{Ko:1987hf}. Thus, the hyperon production in $\bar p$ annihilation
on nuclei is dominated by the interactions of secondary particles produced in $\bar p N$ annihilation
with the nuclear medium. Extensive theoretical calculations within the intranuclear cascade (INC) 
model \cite{Cugnon:1990xw} tend to overestimate the yields of both particle species, 
$\Lambda$ and $K^0_S$, and underestimate the yield ratio $\Lambda/K^0_S$.
The latter might indicate the in-medium enhanced cross sections of the strangeness 
exchange reactions, $\bar K N \to Y \pi$ ($Y$ stands for a $\Lambda$- or $\Sigma$-hyperon).
The enhanced hyperon production has been also interpreted in terms of $\bar p$ 
annihilations on the clusters of nucleons \cite{Cugnon:1984zp,Ko:1987hf} or even the
formation of a cold quark-gluon plasma (QGP)  \cite{Rafelski:1988wn}.

Even more interesting is the double-strange (S=-2) hyperon production in antiproton annihilation on nuclei which
has not yet been studied experimentally. The calculations of Ref. \cite{Ferro:2007zz}, motivated 
by the planned Double-Hypernuclei experiment by PANDA@FAIR \cite{Pochodzalla:2005nf,PANDA}, 
take into account only the direct mechanism $\bar N N \to \bar\Xi \Xi$ which can only account for a few percent
of an inclusive $\Xi$ yield. However, the inclusive production of the 
S=-2 hyperons in $\bar p$ annihilation on nuclei is very interesting by itself. This process 
should be a quite sensitive test for the unusual mechanisms of the $\bar p$ annihilation 
on nuclei, like the QGP formation, since it requires the simultaneous production 
of two $s \bar s$ pairs.

In this work, we study strangeness production in $\bar p$-nucleus collisions at $p_{\rm lab}=0.2-15$ GeV/c 
and in $\bar p$-nucleus annihilation at rest within the microscopic GiBUU transport model 
\cite{Buss:2011mx}. We compare our calculations with all available experimental data 
\cite{Condo:1984ns,Balestra:1987vy,Miyano:1988mq,Ahmad:1997fv}
on $\Lambda$, $K_S^0$ and $\bar\Lambda$ production for in-flight annihilation of $\bar p$ on nuclei
and with earlier calculations within the INC models \cite{Cugnon:1990xw,Strottman:1985jp,Gibbs:1990sf}.
We also analyse the selected data set of ref. \cite{Riedlberger:1989kn} on $\pi^-$, proton
and $\Lambda$ production in $\bar p$ annihilation at rest on $^{14}$N.
In contrast to the INC models, the GiBUU model includes {\it selfconsistent} 
relativistic mean fields acting on baryons, antibaryons, kaons and antikaons.
The selfconsistency means that the mean fields depend on the actual
particle densities and currents. This has never been done in the previous transport 
calculations of antiproton-nucleus reactions. The selfconsistent potential fields are 
very important, for example, for a realistic treatment of annihilation on light nuclei, 
when a nucleus get almost destroyed by pions produced in annihilation, 
and the outgoing particles propagate in much weaker potential fields. 

We study in detail the mechanisms of strangeness production by decomposing spectra and reaction rates 
into the partial contributions from various elementary processes. 
Estimates for the production probabilities of hypernuclei are also given. Finally, we present 
the predictions of our model for the $\Xi$ hyperon spectra. We argue that the latter
can be used to disentangle the hadronic and hypothetic QGP mechanisms of $\bar p$ annihilation
on nuclei.

The structure of the paper is as follows. Sec. \ref{Model} contains a brief description of 
the GiBUU model with an accent on its new and/or improved ingredients, such as, e.g., 
strangeness production channels in $\bar N N$ collisions. In sec. \ref{results}, we present 
the results of our calculations for the $\Lambda$, $K^0_S$ and $\bar\Lambda$ production and compare 
them with experimental data and INC calculations. Then, we give our predictions for 
the $\Xi$ hyperon production at various $\bar p$ beam momenta. 
Finally, sec. \ref{summary} summarizes our work.

\section{Model}
\label{Model}

The GiBUU model \cite{Buss:2011mx} is a transport-theoretical framework which allows to describe
a wide range of photon-, lepton-, hadron-, and nucleus-nucleus reactions.
Below we concentrate mostly on the model ingredients
which govern strangeness production in antiproton-induced reactions. 
For other model details relevant for the present study we refer the reader to refs. 
\cite{Buss:2011mx,Larionov:2008wy,Larionov:2009er,Larionov:2009tc,Gaitanos:2010fd}.

The GiBUU model solves the coupled system of kinetic equations for the different hadronic
species $i=N, \Delta, Y, Y^*, \Xi, \pi, K$ etc and respective antiparticles:
\begin{equation}
  \frac{1}{p_0^*}
  \left[ p^{*\mu} \frac{\partial}{\partial x^\mu} + \left(p_\nu^* F_i^{\mu\nu} 
                                   + m_i^* \frac{\partial m_i^*}{\partial x_\mu}\right)
    \frac{\partial}{\partial p^{* \mu}} \right] f_i(x,{\bf p^*}) 
  = I_i[\{f\}]~,                                                 \label{kinEq}
\end{equation}
where $f_i(x,{\bf p^*})$ is the phase-space distribution function, $p^*=p-V_i$ is the kinetic
four-momentum, $F_i^{\mu\nu} \equiv \partial^\mu V_i^\nu - \partial^\nu V_i^\mu$ is the field tensor,
and $m_i^*=m_i+S_i$ is the effective mass. The scalar potential $S_i=g_{\sigma i}\sigma$
is expressed in terms of the isoscalar $\sigma$ meson ($J^P=0^+$) field. The vector potential
$V_i$ includes the contributions from the isoscalar $\omega$ meson ($J^P=1^-$) field, 
isovector $\vec{\rho}$ meson ($J^P=1^-$) field and the electromagnetic field $A$:
\begin{equation}
   V_i = g_{\omega i} \omega + g_{\rho i} \tau^3 \rho^3 + q A~.            \label{V_i}
\end{equation}
Here, the isovector term $\propto \rho^3$ is included only for nucleons and antinucleons.
The collision term $I_i[\{f\}]$ in the r.h.s. of Eq.~(\ref{kinEq}) represents the contribution 
of binary collisions and resonance decays to the partial time derivative 
$\frac{\partial}{\partial x^0} f_i(x,{\bf p^*})$. The term $\frac{{\bf p^*}}{p_0^*} \nabla f_i(x,{\bf p^*})$ in the l.h.s.
of Eq.~(\ref{kinEq}) is a usual drift term. The term $\propto \frac{\partial}{\partial p^{* \mu}} f_i(x,{\bf p^*})$
describes the deviation of particle trajectories from straight lines as well as 
acceleration/deceleration due to the meson mean
fields. Without this term, the GiBUU model is basically reduced to a usual cascade model.
The meson mean fields $\sigma, \omega, \rho$ and the electromagnetic field $A$ are calculated
from the corresponding Lagrange equations of motion neglecting time derivatives 
\cite{Larionov:2008wy,Gaitanos:2010fd}. This calculation is selconsistent
in the sense that the fields are induced by actual particle densities and currents 
which serve as source terms in the Lagrange equations. 
The full system of transport equations (\ref{kinEq}) and meson field equations 
admits the energy-momentum conservation (c.f. ref. \cite{Larionov:2008wy}).

In order to determine the meson-nucleon
coupling constants and the self-interaction parameters of the $\sigma$ field, 
we use the relativistic mean field (RMF) model
in the NL3 version \cite{Lalazissis:1996rd}.  The coupling
constants of mesons with other nonstrange baryonic resonances ($\Delta, N^*$ etc)
are set equal to the respective meson-nucleon coupling constants. For the
meson-hyperon and meson-kaon coupling constants, we apply a simple light quark counting rule
by putting 
\begin{eqnarray}
   && \frac{g_{\sigma Y}}{2}=g_{\sigma \Xi}=g_{\sigma K}=\frac{g_{\sigma N}}{3}~,  \label{rel1}   \\ 
   && \frac{g_{\omega Y}}{2}=g_{\omega \Xi}=g_{\omega K}=\frac{g_{\omega N}}{3}~.  \label{rel2}
\end{eqnarray}
The coupling constants with the corresponding antiparticles are obtained as follows:
\begin{eqnarray}
   && \frac{g_{\sigma \bar B}}{3\xi}=\frac{g_{\sigma \bar Y}}{2}=g_{\sigma \bar\Xi}=g_{\sigma \bar K}
                                                         =\frac{g_{\sigma N}}{3}~,  \label{rel3} \\
   && \frac{g_{\omega \bar B}}{3\xi}=\frac{g_{\omega \bar Y}}{2}=g_{\omega \bar\Xi}=g_{\omega \bar K}
                                                         =-\frac{g_{\omega N}}{3}~, \label{rel4} \\ 
   && g_{\rho \bar N}=\xi g_{\rho N}~.  \label{rel5}
\end{eqnarray}
Here, ``$B$'' denotes any nonstrange baryon, i.e. $B=N, \Delta, N^*$ etc.
The relative signs in Eqs. (\ref{rel3})-(\ref{rel5}) are obtained from 
the $G$-parities of the meson fields.
The factor $\xi = 0.22$ is introduced in order to obtain the Schr\"odinger
equivalent potential $U_{\bar N} = -150$ MeV for an antiproton at the zero kinetic energy and normal
nuclear density $\rho_0=0.148$ fm$^{-3}$, in agreement with $\bar p$-nucleus scattering 
phenomenology (c.f. \cite{Larionov:2009tc} and refs. therein). One should keep in mind, however, 
that in experiment only the nuclear surface is tested due to a large annihilation cross section.
Thus, an empirical information at $\rho_0$ can only be obtained by model dependent extrapolations. 

In Table \ref{tab:potentials}, we collect the Schr\"odinger equivalent potentials of 
different particles evaluated by using the relations (\ref{rel1})-(\ref{rel5}) and 
the nucleon scalar ($S_N=-380$ MeV) and vector ($V_N^0 = 308$ MeV) potentials 
in nuclear matter at $\rho_0$ \cite{Larionov:2009tc}.
\begin{table}[htb]
\caption{\label{tab:potentials} The Schr\"odinger equivalent potentials of different
particles at zero kinetic energy,\\ $U_i=S_i+V_i^0+(S_i^2-(V_i^0)^2)/2m_i$ (in MeV), 
in nuclear matter at $\rho_0$.}
\begin{ruledtabular}
\begin{tabular}{c|cccccccccc}
 $i$   & $N$  &  $\Lambda$  &  $\Sigma$  &   $\Xi$  & $\bar N$ & $\bar\Lambda$ & $\bar\Sigma$ & $\bar\Xi$ &  $K$  & $\bar K$ \\
\hline
$U_i$  & -46 &  -38      &  -39    &   -22    & -150   & -449     & -449     & -227    & -18  & -224 
\end{tabular}
\end{ruledtabular}
\end{table}
The potential depths for nucleon, $\Lambda$ hyperon and antikaon are consistent with phenomenology
\cite{Friedman:2007zza}. The $\Sigma$ potential is attractive, which contradicts the analysis of
$\Sigma^-$ atoms \cite{Friedman:2007zza}.
A weak attraction for kaons is also not supported by the analysis of the kaon flow from heavy ion 
collisions (c.f. \cite{Larionov:2005kz} and refs. therein), where a weak repulsion has been found. 
These drawbacks are the consequences of a simple treatment of particle potentials based on the 
same RMF model. We do not expect that they sensitively influence our results since the multiplicity 
of $\Sigma$ hyperons is considerably smaller than the multiplicity of $\Lambda$ hyperons while the kaon potential 
is weak anyway. The potentials of the $\Xi$ hyperons and antihyperons are still not restricted by
any experimental data.

The nucleus is modeled by employing a local density approximation.
The momenta of nucleons are sampled uniformly within the spheres of radii
\begin{equation}
   p_{F,i}=(3\pi^2\rho_i)^{1/3}~,~~~i=p,n~,  \label{p_F}
\end{equation}
by using a Monte Carlo method. 
The density profiles of protons and neutrons $\rho_i(r)$ are obtained from a selfconsistent solution of
the relativistic Thomas-Fermi equations \cite{Gaitanos:2010fd}. This makes the nucleus stable on the
time scale of the order of several 100 fm/c, enough for the most reactions with
nuclear targets. The Fermi motion of nucleons results in the smearing of particle production
thresholds, which is especially important for the reactions with low-energy projectiles. 

For the simulation of an $\bar p$-nucleus collision, at $t=0$ an antiproton is placed at the distance 
5 fm + nuclear radius from the nuclear center along the collision axis. 
Such a distant initialization of the antiproton is needed in order to take into account the change of
its momentum and trajectory under the action of attractive nuclear and Coulomb
potentials \cite{Larionov:2009tc}.

For the simulation of annihilation at rest, the initial radial position of the antiproton is
chosen according to the probability distribution (c.f. \cite{Ilinov:1982jk,Cugnon:1986tx})
\begin{equation}
   dP = C |R_{nl}(r)|^2 \rho(r) r^2 dr~.    \label{AnniProb_vs_r}
\end{equation}
where $R_{nl}(r)$ is the radial wave function of the antiproton  in the Coulomb atomic state
with quantum numbers $n$ and $l \leq n-1$, $\rho(r)$ is the nucleus density profile, and $C$ is
a normalization constant.  
Quoting ref. \cite{Cugnon:1986tx}, a cascade of electromagnetic deexcitation of the 
$\bar p$-atom, essentially through the $(n,l=n-1)$ states, emits $X$-rays, which permits
us to trace the $\bar p$ down to the level where annihilation takes place.

Once a nucleus and an antiproton are initialized, the system of kinetic equations (\ref{kinEq})
supplemented by the meson field equations is solved by the test particle method in the parallel
ensemble mode \cite{Bertsch:1988ik}. Between the two-body collisions, the test particle centroids 
$({\bf r},{\bf p}^*)$ propagate according to the Hamiltonian-like equations (c.f. \cite{Larionov:2009tc,Gaitanos:2010fd}).
The two-body collisions are treated in a geometrical picture: the test particles $a$ and $b$ from
the same parallel ensemble are allowed to collide if they pass their minimal distance $d_{\rm min}$ during
a given time step. Here $d_{\rm min} < \sqrt{\sigma_{\rm tot}^{ab}/\pi}$, where $\sigma_{\rm tot}^{ab}$ is the total interaction 
cross section depending on the types of colliding particles and invariant energy $\sqrt{s}$.
In the present work, we use vacuum cross sections neglecting their possible in-medium
modifications. The final state of a two-body collision $ab \to cde...$ is chosen by a Monte Carlo method 
according to the partial cross sections of various outgoing channels. The final state probability
includes the products of Pauli blocking factors $(1-f)$ for the outgoing nucleons.
These factors are calculated from the actual time-dependent neutron and proton phase-space 
distributions.
Collisions between the secondary and primary particles as well as between two secondaries 
are taken into account. The mean fields are recomputed on every time step according 
to the modified particle densities and currents.

\begin{figure}
\includegraphics[scale = 0.80]{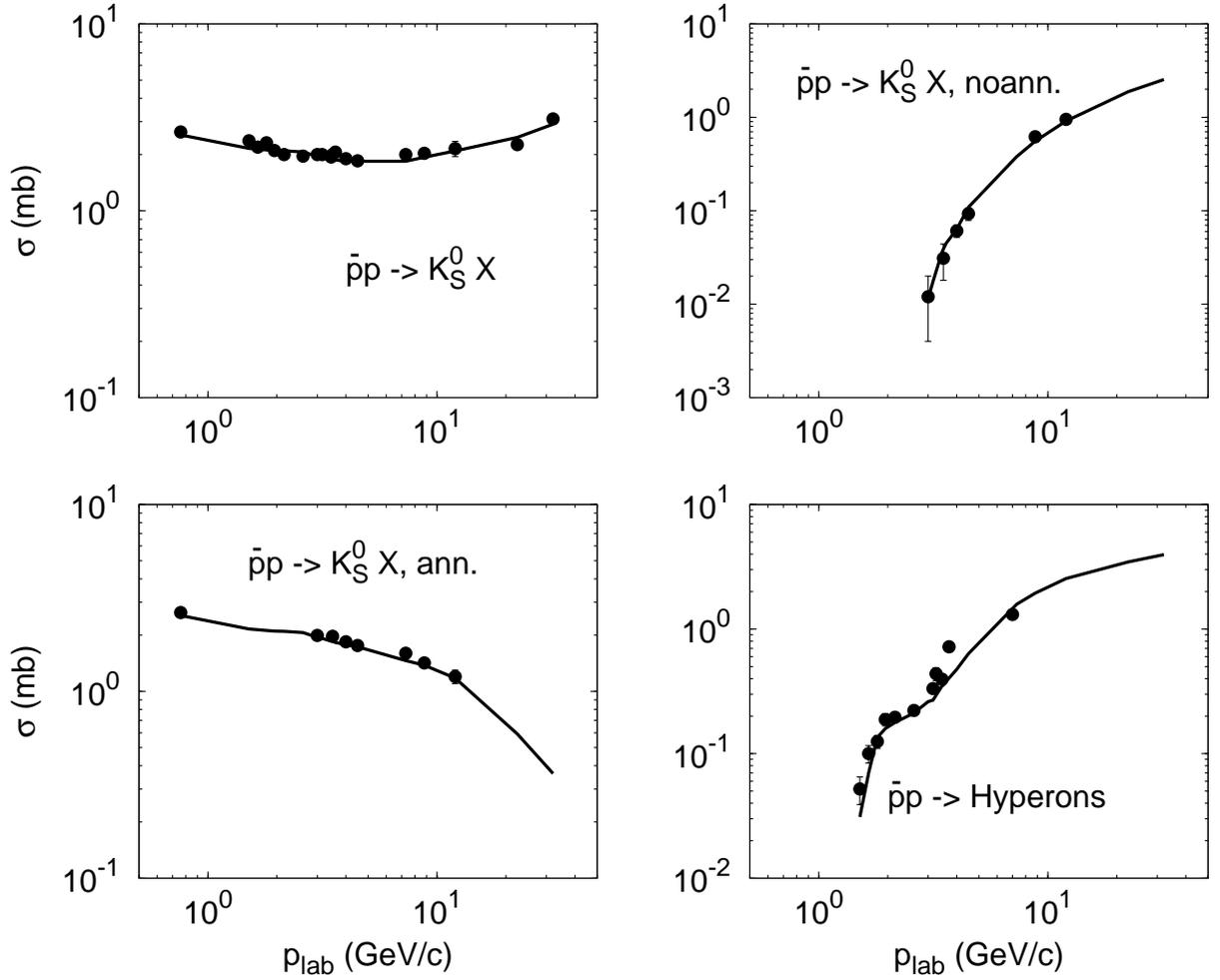}

\vspace*{0.5cm}

\caption{\label{fig:ppbar_to_KS_hyp} Inclusive cross section of $K_S^0$ production
(upper left panel), the cross section of $K_S^0$ production in annihilation
and nonannihilation channels (lower left and upper right panels, respectively),
and the inclusive cross section of the hyperon production (lower right panel)
in antiproton-proton collisions as a function of the beam momentum.
The experimental data are taken from refs. \cite{Ochiai:1984gs,Baldini:1988ti}.}
\end{figure}

In the case of an antinucleon-nucleon collision, the annihilation may result in a large 
number of various mesonic final states. This makes the direct parameterization of all partial 
cross sections practically impossible. Thus, we rely on the statistical annihilation model 
with SU(3) symmetry \cite{PshenichnovPhD,Golubeva:1992tr}.
This model has been successfully applied in the calculations of pion and proton spectra
from $\bar p$ annihilation on nuclei at 608 MeV/c within the GiBUU framework \cite{Larionov:2009tc}.
According to the model \cite{PshenichnovPhD,Golubeva:1992tr}, the probability of 
the $\bar N N$ annihilation to a given final state meson configuration, which may include up to 
$n=6$ mesons $\pi, \eta, \omega, \rho, K, \bar K, K^*$ and $\bar K^*$, is defined as
\begin{eqnarray}
   W_n(\sqrt{s},I_1,...,I_n,Y_1,...,Y_n) &=& w_n(\sqrt{s},I_1,...,I_n,Y_1,...,Y_n)  \nonumber \\
   &\times& a_\pi^{n_\pi} a_\eta^{n_\eta} a_\omega^{n_\omega} a_\rho^{n_\rho} 
            a_K^{n_K+n_{\bar K}} a_{K^*}^{n_{K^*}+n_{\bar K^*}}~,                \label{W_n}
\end{eqnarray}
where $I_1,...,I_n$ and $Y_1,...,Y_n$ are, respectively, the isospins and hypercharges of outgoing mesons,
and $n_i$ are the multiplicities of mesons of each type ($i=\pi, \eta, \omega$ etc).
The quantity $w_n$ is proportional to the phase space volume of a given final state and
is calculated assuming that the incoming and outgoing hadrons can be exactly classified
according to the SU(3) symmetry. The dimensionless parameters $a_\pi, a_\eta,...,a_{K^*}$ break the
exact SU(3) symmetry. They approximate the unknown parts of matrix elements and depend on
the types of the particles and on their internal structure. For the annihilation channels without
strangeness, the values $a_\pi=1, a_\eta=0.13, a_\omega=0.18$ and $a_\rho=0.24$ were determined
in \cite{Golubeva:1994qe} from the best agreement with the data on $\bar p p$ annihilation
at rest and in flight at $p_{\rm lab} \leq 10$ GeV/c \cite{Sedlak:1988vz}. For the parameters
related to the strange mesons, we apply the beam momentum dependent expressions
obtained from the fit of the $\bar p p \to K^0_S X $ and $\bar p n \to K^- K^0_S \pi^+ \pi^-$ 
cross sections (see Fig.~\ref{fig:ppbar_to_KS_hyp} for the $\bar p p$ case):
\begin{equation}
   a_{K(K^*)}=0.07\,(0.05)\,C(p_{\text{lab}})~,               \label{a_K}
\end{equation}
where
\begin{equation}
              C(p_{\text{lab}})=
\cases{
4.2\exp(-1.4p_{\text{lab}})+0.5\exp(-0.16p_{\text{lab}})+1, & for $\bar p p$ \cr
2.0\exp(-0.7p_{\text{lab}})+0.5\exp(-0.16p_{\text{lab}})+1, & for $\bar p n$ \cr
}
\label{Cp}
\end{equation}
with $p_{\text{lab}}$ being the beam momentum (in GeV/c).
The statistical annihilation model works, strictly speaking, only at high beam momenta,
when the particle multiplicities are large. At low beam momenta, this model has to be supplemented 
by the phenomenological branching ratios of the different annihilation channels. For the channels
without strange particles, this has been already done in \cite{PshenichnovPhD,Golubeva:1992tr}. 
In the present work, we have extended the tables of probabilities for various $\bar p p$ and $\bar p n$
annihilation channels at rest \cite{PshenichnovPhD,Golubeva:1992tr} by including the channels 
with strange particles $K \bar K$, $K^* \bar K$ + c.c., and $K^* \bar K^*$ (see Appendix \ref{strProd_atRest}).    
In order to have a smooth transition from these empirical branching ratios to the description
according to the statistical model as the beam energy grows, we determine by a Monte-Carlo
method whether the statistical model itself or the empirical branching ratios are
used to simulate a given $\bar N N$ annihilation event. The probability to choose the tables is
\begin{equation}
   P_{\rm at~rest} = \max\left(0,1-\frac{\sqrt{s}-2m_N}{\sqrt{s_{\rm max}}-2m_N}\right)~,  \label{P_atRest}
\end{equation}
where $\sqrt{s_{\rm max}}=2.6$ GeV is the maximum invariant energy up
to which the annihilation tables at rest still can be selected
(respective beam momentum $p_{\rm lab}=2.5$ GeV/c). At the invariant energies
above $\sqrt{s_{\rm max}}$, the statistical model is used directly. The momenta of outgoing
mesons in an $\bar N N$ annihilation event are always sampled microcanonically according to the 
available phase-space volume, regardless of whether annihilation tables at rest or the statistical
model are applied to choose the flavours and charges of the outgoing mesons. This ensures a smooth
behavior of the kinematics of produced mesons with increasing beam energy. 

Apart from the $\bar N N$ annihilation to mesons, we also include elastic (+charge exchange)
scattering $\bar N N \to \bar N N$, $\Delta$ resonance production
$\bar N N \to \bar N \Delta$ (+c.c.), and hyperon production
$\bar N N \to \bar\Lambda \Lambda,~\bar\Sigma \Lambda (+{\rm c.c}),~\bar\Xi \Xi$ channels.
The corresponding total and angular differential cross section parameterizations 
are obtained from the fits of empirical data.
The cross sections are described in detail in Appendix \ref{elem}.

At $\sqrt{s} > 2.37$ GeV, the inelastic production in $\bar N N$ collisions is simulated with a help
of the \Fritiof{} model \cite{Pi:1992ug}. Exception are the processes $\bar N N \to \bar N N$, 
$\bar N N \to \bar\Lambda \Lambda,~\bar\Sigma \Lambda (+{\rm c.c}),~\bar\Xi \Xi$
which are either not included or not described well in \Fritiof{}.
Thus, we treat these processes separately according to their partial cross sections
at any $\sqrt{s}$. 

Figure \ref{fig:ppbar_to_KS_hyp} shows the inclusive cross sections of $K_S^0$
and hyperon production in $\bar p p$ collisions. The partial cross sections of $K_S^0$ 
production in the annihilation and nonannihilation events are also shown 
in the same figure. The experimental data on $K_S^0$ production are described very well,
including the nonannihilation channels, which are simulated by the \Fritiof{} model.
The hyperon production cross section is somewhat underestimated at the beam 
momenta $\sim 4$ GeV/c. This is mainly due to still underestimated cross section of the
$\bar p p \to \bar\Lambda \Lambda X$ inclusive channel above 3 GeV/c, which is practically missed
in the \Fritiof{} model. However, we find that, overall, the \Fritiof{} model describes 
the inelastic production in $\bar p p$ collisions better, than, e.g.,
the \Pythia{} model \cite{Sjostrand:2006za}. Generally, the latter is successfully 
employed in GiBUU for the description of baryon-baryon and meson-baryon collisions at high invariant 
energies ($\sqrt{s} >  2.6$ GeV and $\sqrt{s} >  2.2$ GeV, respectively). We have to only
admit one problem. The \Pythia{} model does not include $K^0$ and $\bar K^0$ in its list 
of the possible incoming particles. Thus, both of them are replaced by $K^0_L$ in GiBUU 
every time when \Pythia{} is used. This leads to $\sim 5\%$ violation of the total 
strangeness conservation in our calculations, which, however, is accurate enough
for the present exploratory studies. 

The processes of a hyperon scattering, $\Lambda N \to \Lambda N$, 
$\Lambda N \leftrightarrow \Sigma N$, $\Sigma N \to \Sigma N$, $\Xi N \to \Xi N$, $\Xi N \to \Lambda \Lambda$,
$\Xi N \to \Lambda \Sigma$, and of the strangeness exchange on nucleons, $\bar K N \to Y \pi$,
$\bar K N \to \Xi K$, are taken into account in the model. When present, their 
empirical cross sections have been suitably parameterized or the fits
to the existing theoretical calculations have been done \cite{Gaitanos:2011fy,effe_phd}.
As we will see later on, the strangeness exchange processes are very important 
for the hyperon production in antiproton-induced reactions on nuclei. 
In GiBUU, the strangeness exchange reactions $\bar K N \to Y \pi$ are partly 
mediated by the $S=-1$ hyperon resonance formations and decays $\bar K N \to Y^* \to Y \pi$
\cite{effe_phd}.

The associated hyperon production is included in GiBUU via reaction
channels $\pi N \to Y K$, $\eta N \to Y K$, $\rho N \to Y K$ and $\omega N \to Y K$. The cross
sections $\pi N \to Y K$ have been parameterized according to Ref. \cite{Tsushima:1996tv}.
The $\eta$-induced associated hyperon production cross
sections on proton have been reconstructed from the respective cross sections of the 
$\pi^0$-induced processes at the same invariant energy $\sqrt{s}$ by utilizing 
the detailed balance relations \cite{Cugnon:1989rj} and isospin invariance:
\begin{eqnarray}
   \sigma(\eta p \to K^+ \Lambda)  &=& \sigma(\pi^0 p \to K^+ \Lambda) 
                                       \frac{p_{\pi N}}{p_{\eta N}}~,  \\
   \sigma(\eta p \to K^+ \Sigma^0) &=& \sigma(\pi^0 p \to K^+ \Sigma^0) 
                                       \frac{p_{\pi N}}{p_{\eta N}}~, \\
   \sigma(\eta p \to K^0 \Sigma^+) &=& 2 \sigma(\eta p \to K^+ \Sigma^0),~ 
\end{eqnarray}
where $p_{\pi N}$ and $p_{\eta N}$ are the center-of-mass (c.m.) momenta for 
the corresponding initial channels calculated at the same $\sqrt{s}$. 
Similar formulas have also been used for the $\omega p$ initial channel. 
For the $\rho$-induced reactions on proton, we have assumed
\begin{eqnarray}
   \sigma(\rho p \to K \Lambda)  &=& \sigma(\pi p \to K \Lambda)
                                     \frac{p_{\pi N}}{p_{\rho N}}~,  \\
   \sigma(\rho p \to K \Sigma)  &=& \sigma(\pi p \to K \Sigma)
                                     \frac{p_{\pi N}}{p_{\rho N}}~,
\end{eqnarray}
where the isospin states of all particles match each other.
The cross sections of the $\eta$-, $\omega$- and $\rho$-induced reactions
on neutron have been obtained by using the isospin reflection from
the corresponding cross sections on proton.

We note, finally, that the bubble chamber data on $\Lambda$ production 
contain also the admixture of $\Lambda$'s produced by the decays $\Sigma^0 \to \Lambda \gamma$.
These decays are not included in GiBUU, since the $\Sigma^0$ life time 
$\tau=7.4 \times 10^{-20}~{\rm s} \simeq 222000~{\rm fm/c}$ is much longer than the typical hadron-nucleus 
reaction time scale of $\sim 100$ fm/c. Thus, in the present calculations, 
we simply add the $\Sigma^0$ yield to the $\Lambda$ yield.

\section{Results}
\label{results}

\subsection{Time evolution of hyperon production}

In our calculations, baryons experience the action of attractive mean 
field potentials. Slow hyperons get captured inside the residual excited
nuclear system. This system may evaporate particles and/or decay into
fragments, some of which will be single- or double-$\Lambda$ hypernuclei.
It is natural to assume that the fragmentation and evaporation will not change 
much the total yield of hypernuclei; they may, however, affect the production
of a given hypernuclear species. 

In order to distinguish the hyperons outside and inside the residual nucleus, 
we have applied a simple criterion based on the relative distance between particles 
\cite{Aichelin:1991xy}: a particle is outside the nucleus if it is separated by the 
distance larger than some critical distance $d_c$ from all other particles 
of the nucleus. Otherwise, the particle is inside the nucleus. 
Provided the evolution time is long enough, the result should not be much
influenced by the choice of $d_c$, if the latter is larger than the internucleon 
spacing $\sim 1-2$ fm.
\begin{figure}
\includegraphics[scale = 0.80]{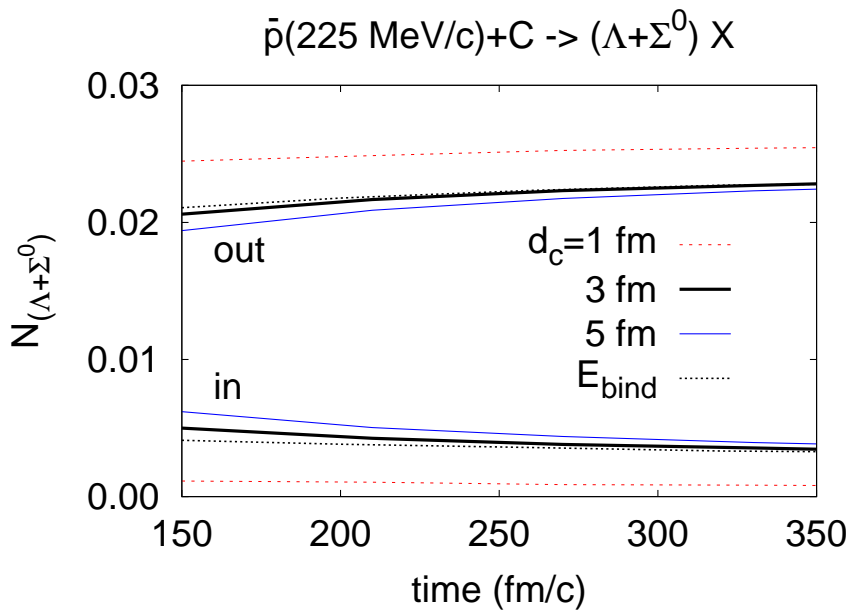}

\vspace*{0.5cm}

\caption{\label{fig:sig_Lambda_vs_time_dc}
The number of $(\Lambda+\Sigma^0)$ hyperons outside (out) and inside (in) 
the residual nucleus per annihilation event as a function of time
for $\bar p$ annihilations at 225 MeV/c on $^{12}$C.
The calculations are done for three different critical
distances $d_c$ and for the criterion according to the hyperon binding energy.
The results are weighted with the impact parameter of 
an incoming $\bar p$.}
\end{figure}
This is illustrated in Fig.~\ref{fig:sig_Lambda_vs_time_dc} where we show the number of 
$(\Lambda+\Sigma^0)$ hyperons outside and inside the residual nucleus per one annihilation 
event as a function of time for $\bar p$(225 MeV/c)+$^{12}$C collisions for various choices of
the critical distance. After $\sim 200$ fm/c the both numbers change
very slowly indicating the presence of really captured $\Lambda$ hyperons
inside the attractive potential well. For comparison, we also present the results
for the bound and unbound hyperons in the same figure. In this case a hyperon $Y$ 
is considered to be bound (unbound) if $E_Y < m_Y$ ($E_Y > m_Y$), where
$E_Y=\sqrt{({\bf p^*})^2+(m_Y^*)^2}+V_Y^0$ is its single-particle energy. This criterion allows
to identify the captured hyperons somewhat earlier. However, after 200 fm/c 
it is very close to the criterion according to $d_c=3$ fm.

Of course, the capture may only happen if a particle experiences the 
action of an attractive potential.
\begin{figure}
\includegraphics[scale = 0.80]{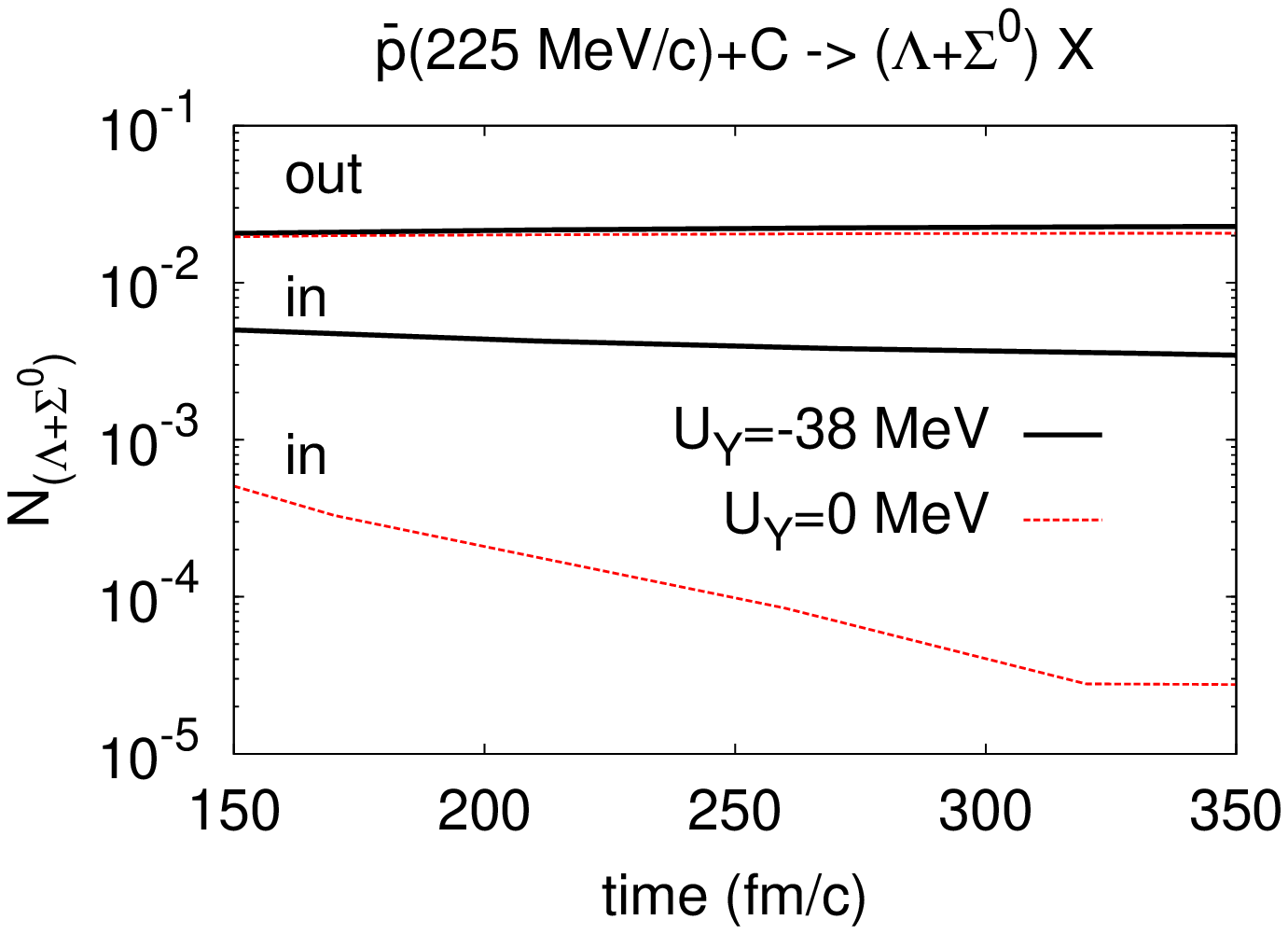}

\vspace*{0.5cm}

\caption{\label{fig:sig_Lambda_vs_time_U}
Same as in Fig.~\ref{fig:sig_Lambda_vs_time_dc} but for calculations
with and without $\Lambda$-potential at the fixed critical distance $d_c=3$ fm.}
\end{figure}
Fig.~\ref{fig:sig_Lambda_vs_time_U} compares the time dependence of the number of
$\Lambda$'s inside and outside the nucleus for calculations with and 
without $\Lambda$-potential. As expected, in the calculations without 
$\Lambda$-potential, the number of hyperons inside the nucleus quickly drops 
with time indicating that there are no captured $\Lambda$'s in this case.  

In the following, we always fix $d_c=3$ fm as in ref. \cite{Aichelin:1991xy}
and calculate the time evolution until 200 fm/c.
Our results for the number of free hyperons, i.e. those outside the residual nucleus,
change only by $\sim 10\%$ if we further increase the evolution time
(c.f. Fig.~\ref{fig:sig_Lambda_vs_time_dc}). 
\begin{figure}
\includegraphics[scale = 0.80]{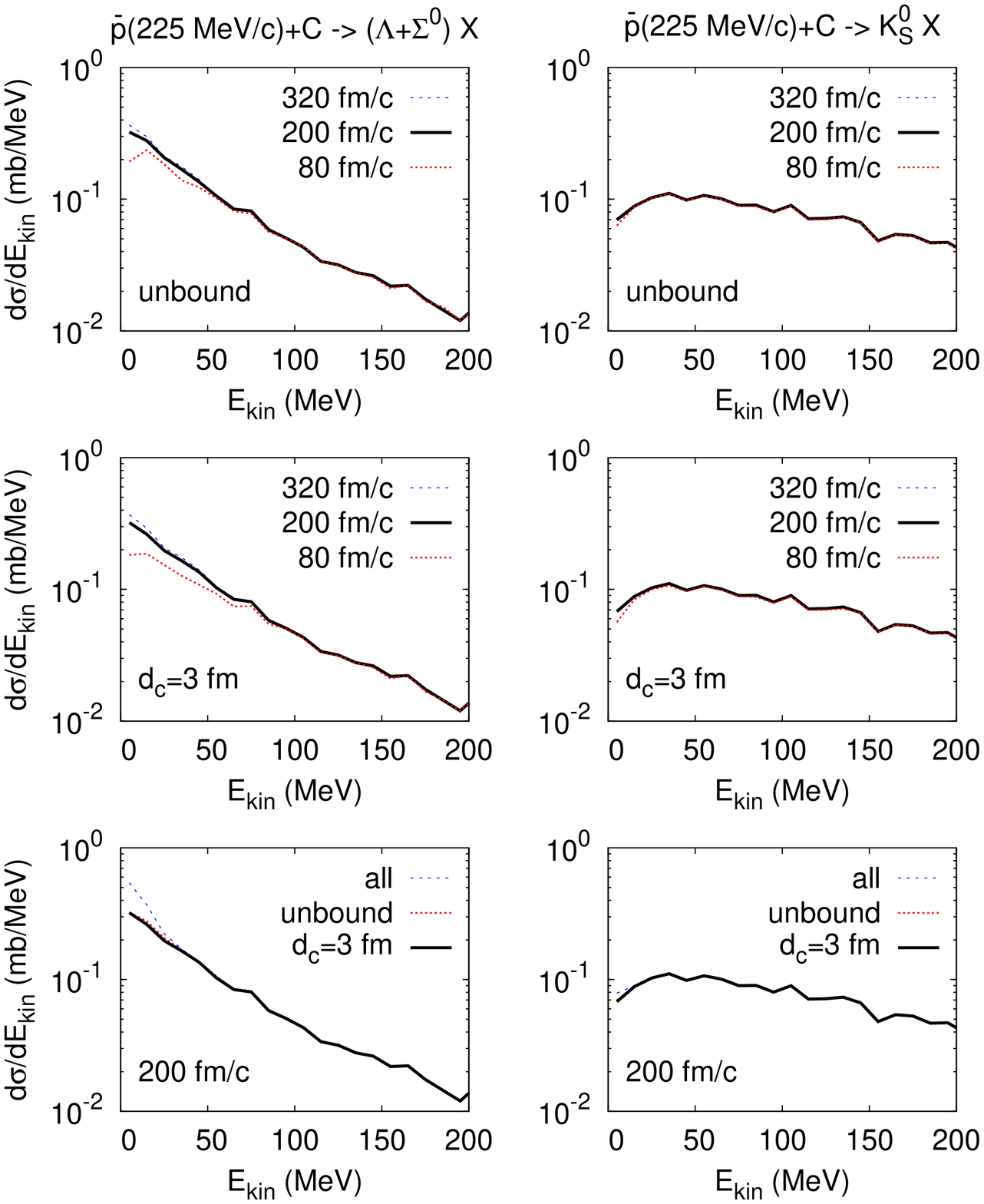}
\caption{\label{fig:dsig_dE_tdep} Kinetic energy spectra of 
$(\Lambda+\Sigma^0)$ hyperons (left panels) and $K^0_S$ mesons (right panels).
The spectra of free particles selected according to the two different criteria,
$E_i > m_i$ ($i=Y, K$, upper panels) and  $d_c=3$ fm (middle panels), are plotted at 
different times. Also the comparison of the spectra obtained with these two
criteria with the spectrum of all particles, i.e., without any restrictions 
is shown at 200 fm/c (lower panels). The number of $K_S^0$ mesons has been determined
from the numbers of $K^0$ and $\bar K^0$ as $N_{K_S^0}=\frac{1}{2}(N_{K^0}+N_{\bar K^0})$.}
\end{figure}
As demonstrated in Fig.~\ref{fig:dsig_dE_tdep}, this change concerns only slow hyperons,
while the yields of fast particles are practically stable.
In particular, the kaon yields and spectra are not influenced
by any further increase of the evolution time.

\subsection{Annihilation at rest}

We selected the ASTERIX@LEAR data \cite{Riedlberger:1989kn} on charged pion, proton and $\Lambda$ production
from $\bar p$ annihilation at rest on $^{14}$N. According to ref. \cite{Poth:1977kg}, the last 
observable transition in light antiprotonic atoms is  $4 \to 3$. Thus we assumed that the antiproton 
occupies mainly the $n=3,l=2$ level immediately before annihilation which we used as an
input in our calculations (see Eq. (\ref{AnniProb_vs_r})). We have checked, however,
that our results are changed by only few percent if the quantum numbers $n=4,l=3$ are chosen for 
the antiproton wave function.

\begin{figure}
\includegraphics[scale = 0.80]{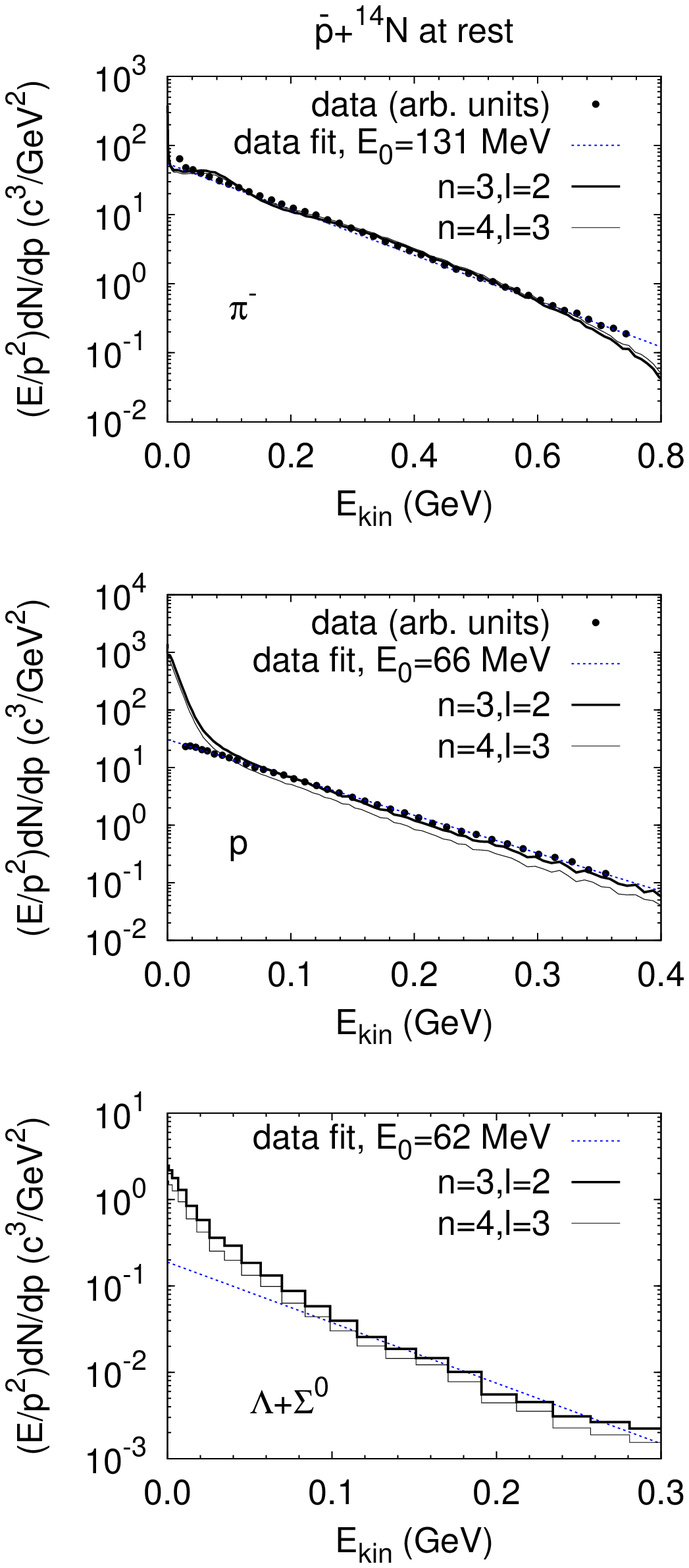}
\caption{\label{fig:dsig_dE_pbarN_atrest} Kinetic energy spectra of negative pions, protons
and $(\Lambda+\Sigma^0)$ hyperons for the $\bar p$ annihilation at rest on $^{14}$N.
The GiBUU calculations are performed for the n=3, l=2 and n=4, l=3 states
of the antiproton.
Experimental data and their fits according to the relativistic
Maxwell-Boltzmann distribution of Eq. (\ref{relMB}) are from ref. 
\cite{Riedlberger:1989kn}. The slope temperatures $E_0$ for the experimental
spectra are also given on the plots.}
\end{figure}
Fig.~\ref{fig:dsig_dE_pbarN_atrest} shows the $\pi^-$, $p$ and $\Lambda$ kinetic energy spectra 
in comparison with our calculations. In ref. \cite{Riedlberger:1989kn}, the measured spectra 
were fitted by the relativistic Maxwell-Boltzmann distribution 
\begin{equation}
    E\frac{dN}{p^2dp} = A \exp(-E/E_0)~,    \label{relMB}
\end{equation}
with the normalization $A$ and the temperature $E_0$ being fit parameters.
We observe that the high energy parts of calculated spectra --- except for
very high energy ($E_{\text{kin}} > 0.6$ GeV) pion spectrum --- agree with 
the data and with the Maxwell-Boltzmann formula (\ref{relMB}) reasonably well. 
A slight underprediction of the high-energy pion spectrum can be traced back
to the case of $\bar p p$ annihilation annihilation at rest 
(c.f. Fig.~B.52 in ref. \cite{Buss:2011mx}).
More significant is the deviation from the data at small kinetic energies.
We predict the strongly enhanced evaporative emission of the low-energetic particles, 
in-particular, $p$ and $\Lambda$, in contrast with the ASTERIX data.
The enhanced emission of low-energy nucleons is also present in the INC calculations of 
ref. \cite{Cugnon:1986tx} for the $\bar p$ annihilations at rest on $^{98}$Mo.  

The calculated pion spectrum also has a clear two-component structure, which seems to be absent
in the pion spectra measured by ASTERIX. The high energy pions ($E_{\rm kin} > 300$ MeV or $p_{\rm lab} > 400$ MeV/c) 
originate directly from $\bar N N$ annihilation almost without rescattering on nucleons. The low energy pions 
are mostly the products of the $\pi N \to \Delta \to \pi N$ processes. This structure is present in the CALLIOPE@LEAR 
data \cite{Mcgaughey:1986kz} for the pion momentum spectra from 608 MeV/c antiproton annihilation 
on $^{12}$C and $^{238}$U which agree with the GiBUU calculations very well \cite{Larionov:2009tc}.

Although the radial distributions of annihilation points are somewhat different for
annihilation at rest and low-energy annihilation in-flight, the angle-integrated 
momentum spectra of emitted particles are expected to change only a little 
\cite{Cugnon:1986tx}. Therefore, we attribute the above discrepancies to the 
reduced acceptance of the ASTERIX spectrometer for the low-momentum particles, 
as also mentioned by the authors themselves in ref. \cite{Riedlberger:1989kn}.

\subsection{Annihilation in-flight}

\begin{figure}
\includegraphics[scale = 0.80]{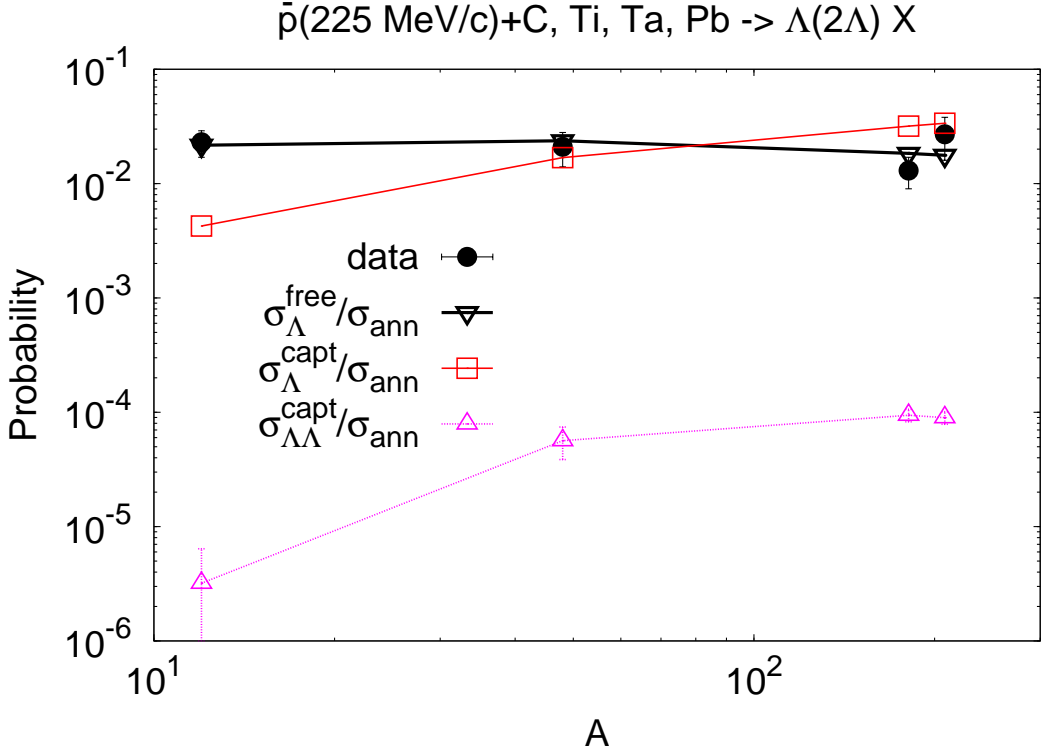}

\vspace*{0.5cm}

\caption{\label{fig:sig_Lambda_vs_A}
The probability of a free $\Lambda$ as well as one and two captured $\Lambda$'s
production per annihilation event vs the target mass number for annihilations
of $\bar p$ at $p_{\rm lab}=225$ MeV/c on $^{12}$C,$^{48}$Ti,$^{181}$Ta 
and $^{208}$Pb. The experimental data for free $\Lambda$ are from 
\cite{Condo:1984ns}.}
\end{figure}
We start from the lowest beam momenta and consider the reactions
$\bar p$(0-450 MeV/c)+$^{12}$C,$^{48}$Ti,$^{181}$Ta and $^{208}$Pb
measured at BNL \cite{Condo:1984ns}.
Figure \ref{fig:sig_Lambda_vs_A} shows the calculated target mass 
dependence of the free $\Lambda$ production probability per 
annihilation event together with the data. 
Note, that the $\bar p$-annihilation cross section $\sigma_{\rm ann} \propto A^{2/3}$, i.e., 
it grows with the target mass number. However, we got rid of this enhancement
by dividing out $\sigma_{\rm ann}$ from the $\Lambda$ production cross sections. 
We also present the calculated production probabilities of nuclear
systems with one and two captured $\Lambda$ hyperons in the same
reactions. In calculations, the beam momentum was fixed at 225 MeV/c.
The free $\Lambda$ production probability is weakly sensitive to the
target mass number and agrees with experiment. However, the probabilities
of one and two captured $\Lambda$ production grow with
the target mass number by almost one order of magnitude from the lightest
($^{12}$C) to the heaviest ($^{208}$Pb) target. This is expected
since, in heavier targets, the produced hyperons are more efficiently 
decelerated by rescattering on nucleons. The detailed calculations of
hyperfragment production in $\bar p$ annihilation on nuclei within the
coupled GiBUU + statistical multifragmentation models are presented in 
Ref. \cite{Gaitanos:2011fy}.
 
\begin{figure}
\includegraphics[scale = 0.60]{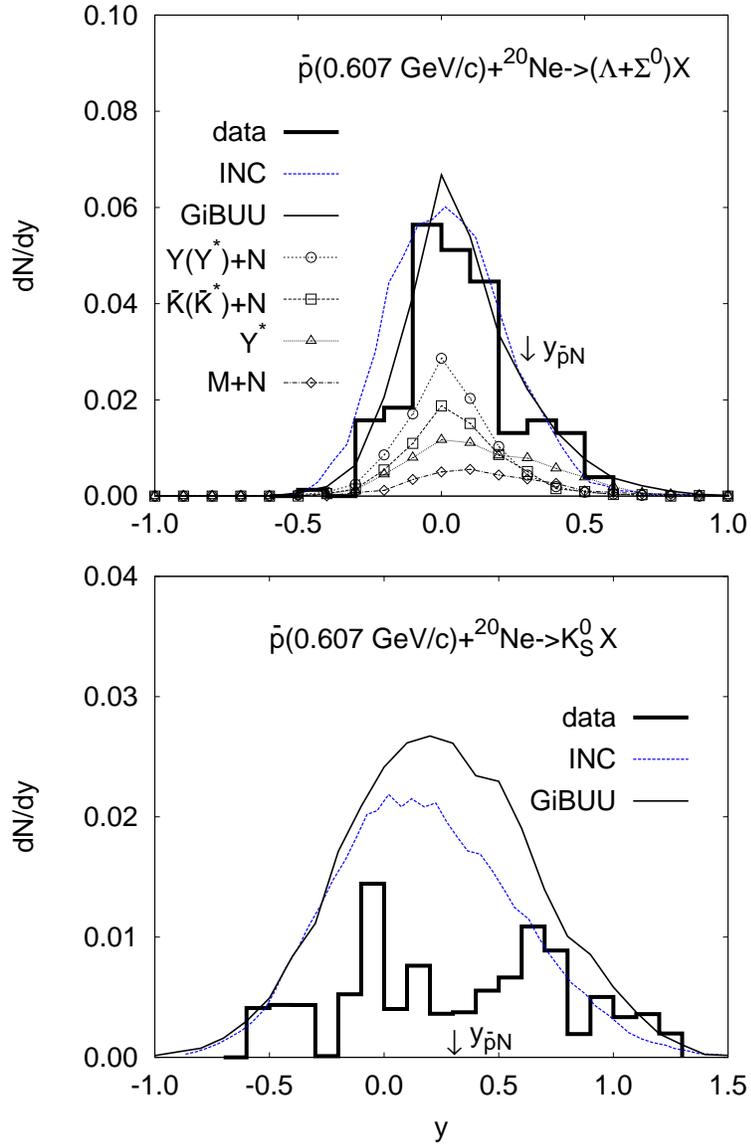}

\vspace*{0.5cm}

\caption{\label{fig:dsig_dy_pbarNe}
The summed rapidity distributions of $\Lambda$ and $\Sigma^0$ hyperons
(upper panel) and the rapidity distribution of $K_S^0$ mesons (lower panel) produced in $\bar p$ 
annihilations on $^{20}$Ne at 607 MeV/c --- thin solid lines. The integrated distributions
give the number of particles per annihilation event.
The partial contributions to the hyperon rapidity spectrum 
(``M'' means any nonstrange meson $\pi, \eta, \rho, \omega$) are
depicted by lines with open symbols.
The INC calculations from ref. \cite{Cugnon:1990xw}
are shown by dashed lines. The experimental data (histograms) are taken 
from ref. \cite{Balestra:1987vy}. The c.m. rapidity of an antiproton and
a nucleon $y_{\bar p N}=0.305$ is shown by arrows.}
\end{figure}
Figure~\ref{fig:dsig_dy_pbarNe} shows the rapidity distributions of 
$\Lambda$ hyperons and $K_S^0$ mesons from $\bar p$+$^{20}$Ne collisions at
608 MeV/c in comparison to the data and to the INC calculations
from \cite{Cugnon:1990xw}.
As we observe, both models agree within $\sim 30\%$. The experimental $\Lambda$ rapidity 
distribution is rather well described, while the calculated $K_S^0$ spectra
are clearly above the data. Since the direct channel of hyperon
production, $\bar p p \to \bar \Lambda \Lambda$ ($p_{\rm lab}^{\rm thr}=1.439$ GeV/c), is closed at 
$p_{\rm lab}=608$ MeV/c, the hyperons can only be produced in strangeness 
exchange processes $\bar K (\bar K^*) N \to Y (Y^*) \pi$, in hyperon resonance 
formation reactions $\bar K N \to Y^*$, or in meson-nucleon collisions 
($\omega N \to Y K$ mostly). 
Our calculation produces somewhat more $K_S^0$'s than the INC calculation
\cite{Cugnon:1990xw} does. This is partly due to taking into account the
target destruction by $\bar p$ annihilation in our calculations, which reduces
the chances for an $\bar K$ to be absorbed. There is also another possible
reason for the different results. Both models contain the $\bar K N$ 
elementary cross sections fitted to the experimental data.
However, the important difference is that in GiBUU the $Y^*$ resonances
have finite life times $\sim 1/\Gamma$ and, therefore, can escape out of the
nucleus. This effectively reduces the $\bar K$ absorption since a $Y^*$
excited as the intermediate state of $\bar K N$ scattering, $\bar K N \to Y^* \to \bar K N$,
``hides'' an $\bar K$ from interactions with other nucleons.

The $(\Lambda+\Sigma^0)$ rapidity spectrum shown in Fig.~\ref{fig:dsig_dy_pbarNe}
is decomposed to its partial components according to various elementary production
channels of $\Lambda$ or $\Sigma^0$. We see that the largest contribution is given by
the hyperon rescattering on nucleons with flavour and/or charge exchange,
in-particular, by the exothermic processes $\Sigma^+ n \to \Lambda p$ and $\Sigma^- p \to \Lambda n$.
The main channel of the hyperon production, however, is the strangeness exchange
processes $\bar K (\bar K^*) N \to Y(Y^*) \pi$. The rescattering $\Lambda(\Sigma_0) N \to  \Lambda(\Sigma_0) N$
somewhat distorts the shapes of these partial contributions by shifting the produced
$\Lambda$'s($\Sigma ^0$'s) towards smaller absolute values of rapidity 
(see also Fig.~\ref{fig:dsig_dy_pbarTa_Lambda} below).

A more detailed information on the relative
importance of various hyperon production channels is given in Table~\ref{tab:pbarNe}. 
\begin{table}[htb]
\caption{\label{tab:pbarNe} Partial contributions of the 
different reactions to the multiplicities of $\Lambda$, $\Sigma$
$\bar K$ and $K$ per 1000 $\bar p$ absorption events on $^{20}$Ne
at 608 MeV/c. In the calculation of reaction rates we did
not distinguish kaons from $K^*$'s and antikaons
from $\bar K^*$'s. The negative numbers correspond to the absorption
reaction channels for a given particle. Results of the INC calculations 
\cite{Cugnon:1990xw} are given in brackets.}

\vspace*{1cm}

\begin{tabular}{llllllll}
\hline
$\bar B B \to \Lambda K$     & 0 (0) &
$\bar B B \to \Sigma K$     & 0 (0) &
$\bar B B \to \bar K X$      & 57.7 (50.0) &
$\bar B B \to K X$         & 57.7 (50.0)    \\
$\pi B \to \Lambda K$        & 0.6 (1.4) &
$\pi B \to \Sigma K$        & 0.6 (1.0) & 
$\Sigma M \to \bar K X$      & 0.2       &
$\phi \to K \bar K$        & 1.9       \\
$\eta B \to \Lambda K$      & 0.2 (1.5) &
$\eta B \to \Sigma K$       & 0.3 (0.4) &
$Y^* M \to \bar K X$        & 0.5   &
$\pi B \to \Lambda K$       & 0.6 (1.4)   \\
$\rho B \to \Lambda K$      & 0.9  &
$\rho B \to \Sigma K$       & 1.5  &
$\phi \to K \bar K$        & 1.9 &
$\eta B \to \Lambda K$      & 0.2 (1.5)   \\
$\omega B \to \Lambda K$     & 1.5 (5.0) &
$\omega B \to \Sigma K$      & 3.1 (4.0) &
$ M M \to K \bar K$        & 3.2   &
$\rho B \to \Lambda K$      & 0.9         \\
$\bar K B \to \Lambda X$     & 8.4 (10.0) &
$\bar K B \to \Sigma X$      & 10.7 (16.0) &
$K \bar K \to M M$         & -1.2  &
$\omega B \to \Lambda K$     & 1.5 (5.0)     \\
$\Sigma B \to \Lambda X$     & 9.9 (10.0) &
$\Sigma B \to \Lambda X$     & -9.9 (-10.0) &
$\bar K B \to \Lambda X$     & -8.4 (-10.0) &
$\pi B \to \Sigma K$        & 0.6 (1.0)    \\
$Y^* B \to \Lambda X$       & 1.3 &
$Y^* B \to \Sigma X$        & 1.4   &
$\bar K B \to \Sigma X$       & -10.7 (-16.0) &
$\eta B \to \Sigma K$        & 0.3 (0.4)    \\
$\Lambda B \to \Sigma X$      & -0.9 (-0.9) &
$\Lambda B \to \Sigma X$      & 0.9 (0.9) &
$\bar K B \to Y^*$          & -17.0 &
$\rho B \to \Sigma K$        & 1.5    \\
$\Lambda M \to Y^* X$        & -1.5 &
$\Sigma M \to Y^* X$         & -1.5 &
$Y^* \to \bar K N$         & 6.4 &
$\omega B \to \Sigma K$      & 3.1 (4.0)     \\
$Y^* \to \Lambda \pi(\eta)$    & 5.9 &
$Y^* \to \Sigma \pi$        & 7.6 &
$\bar K B \to Y^* \pi$       & -2.7 &
$ M M \to K \bar K$         & 3.2    \\
$BB \to \Lambda X$           & 0.2 &
$\Sigma M \to \bar K X$      & -0.2 & 
                     &     &
$K \bar K \to M M$         & -1.2  \\
                     &     &
                     &     &
                     &     &
$B B \to K X$            & 0.3 \\
Sum $\Lambda$          & 26.5 (27.0) &
Sum $\Sigma$           & 14.5 (12.3) &
Sum $\bar K$           & 29.9 (24.0) &
Sum $K$              & 70.6 (63.3)  \\
\hline
\end{tabular}
\end{table}
As one can see by inspecting Table~\ref{tab:pbarNe}, $\sim 80\%$ of the total $Y$ and $Y^*$
production rate (without $Y$ and $Y^*$ absorption contributions) is caused by the 
$\bar K B \to Y X$, $\bar K B \to Y^*$ and $\bar K B \to Y^* \pi$ 
secondary processes. Nevertheless, the associated production $M B \to Y K$ constitutes the remaining 
$20\%$ and, therefore, is quite important as well. In contrast, 80\% of the kaon and antikaon production 
rate is caused by the direct mechanism $\bar B B \to K \bar K X$. 
The INC calculation gives larger contributions of the antikaon absorption $\bar K B \to Y X$ and 
associated production $M B \to Y K$. At the same time, the hyperonic decays of $Y^*$ resonances present in
GiBUU counterbalance the smaller $\Lambda$ and $\Sigma$ hyperon production by other channels. This leads
to rather similar final results for the both models.

It is quite interesting to observe from Table~\ref{tab:pbarNe} that the $\pi B \to Y K$ channels make
relatively small contributions to the $\Lambda$ and $\Sigma$ production rates with respect to the
$\omega B \to Y K$ channels even though there is an abundant pion production in $\bar p$ annihilation
on nuclei.
The reason is that in $\bar p$ annihilation at rest or at low $\bar p$ beam momentum  
most of pions are produced with momenta $\sim 0.2-0.4$ GeV/c (c.f. Fig.~2 in 
ref. \cite{Larionov:2009tc}, where the $\pi^+$ momentum spectrum is shown for 
$\bar p$+$^{12}$C collisions at 608 MeV/c). This is well below the threshold
pion momentum of 0.895 GeV/c for the $\pi N \to \Lambda K$ reaction on a nucleon
at rest. On the other hand, an $\omega$ meson is produced in a rather large 
fraction ($\sim 20\%$) of $\bar N N$ annihilation events at rest 
(c.f. ref. \cite{Golubeva:1992tr}). This favors the exothermic
$\omega N \to Y K$ reactions, which have a large cross section at low $\omega$ momentum.
The situation is, however, different for the energetic $\bar p$-nucleus collisions 
(see Table~\ref{tab:pbarTa} below). 

\begin{figure}
\includegraphics[scale = 0.70]{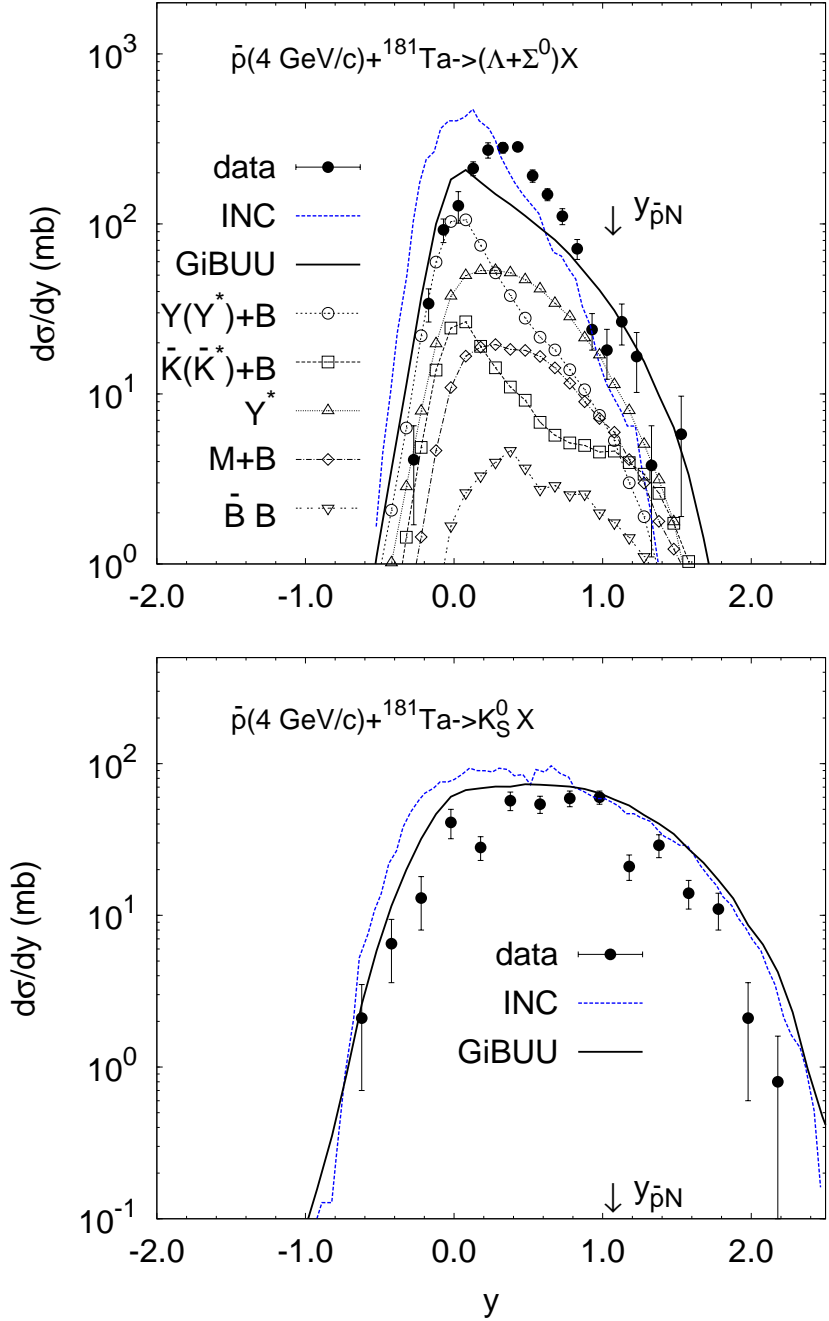}
\caption{\label{fig:dsig_dy_pbarTa_decomp}
The summed rapidity distributions of $\Lambda$ and $\Sigma^0$ (upper panel)
and the rapidity distribution of $K^0_S$ (lower panel) for the
collisions $\bar p$(4 GeV/c)+$^{181}$Ta are shown by solid lines.  
The partial contributions to the hyperon rapidity spectrum 
(``M'' means any nonstrange meson $\pi, \eta, \rho, \omega$,
``B'' --- any nonstrange baryon $N, \Delta, N^*$ etc.) are
shown by various lines with open symbols.
The INC calculations (dashed lines) are taken from ref. \cite{Cugnon:1990xw}.
The experimental data (solid circles) are from ref. \cite{Miyano:1988mq}.
The antiproton-nucleon c.m. rapidity $y_{\bar p N}=1.078$ is shown by vertical arrows.}
\end{figure}
At 4 GeV/c, the agreement of our calculations with the experimental $K_S^0$ spectrum becomes better, 
as one can see from Fig.~\ref{fig:dsig_dy_pbarTa_decomp}. However, the ($\Lambda+\Sigma^0$) yield
is now underpredicted at intermediate rapidities $y \simeq 0.5 y_{\bar p N}$, where $y_{\bar p N}$
is the rapidity of an antiproton-nucleon c.m. system. From the spectrum decomposition
into partial channels, which is also given in Fig.~\ref{fig:dsig_dy_pbarTa_decomp}, we
observe qualitatively the same picture as at lower beam momentum 607 MeV/c (Fig.~\ref{fig:dsig_dy_pbarNe}
above). Some changes with respect to the lower beam momentum are, however, visible: 
The hyperon resonance $Y^*$ decays grow to the second place in importance. 
The relative contribution of nonstrange meson-baryon collisions is increased and becomes as 
important as that of strangeness-exchange reactions. 
Also a nonnegligible contribution of direct $\bar B B$ channels appears now.

The disagreement with the experimental rapidity spectrum of ($\Lambda+\Sigma^0$) hyperons at
$y \simeq 0.5$ might be due to still underpredicted contribution of the $Y^* \to Y \pi$ decays.
Indeed, this contribution has a broad 
maximum at $y \simeq 0.3$, i.e., close to the maximum of the measured rapidity spectrum. 
Since our calculations tend to overestimate the $K_S^0$ production, we can assume that the
$\bar K N \to Y^*$ cross sections should be larger. In-particular, the total cross section 
$K^- n \to X$ (or the same cross section of the isospin-reflected channel $\bar K^0 p \to X$) 
is underestimated at $\sqrt{s} < 1.7$ GeV (see Fig. 2.15 in ref. \cite{effe_phd}). 

Another reason for the deviation with the experimental rapidity spectrum of
($\Lambda+\Sigma^0$) hyperons is related to rather 
uncertain $Y N \to Y N$ cross sections. Especially the $\Lambda N \to \Lambda N$ cross section
is important, as has been noticed, first, by Gibbs and Kruk in ref. \cite{Gibbs:1990sf}.
These authors have applied their INC code \cite{Strottman:1985jp,Gibbs:1990sf} to the reaction 
$\bar p$(4 GeV/c)+$^{181}$Ta and reproduced the measured $(\Lambda+\Sigma^0)$ and $K_S^0$
rapidity spectra very well. They assigned a constant elastic hyperon-nucleon cross
section of 14 mb, which is about two times less than the $\Lambda p$ elastic cross section 
at $p_{\rm lab}\simeq0.5$ GeV/c (corresponding to the peak position $y\simeq0.4$
of the measured $\Lambda$ rapidity spectrum, c.f. Fig.~\ref{fig:dsig_dy_pbarTa_decomp}) 
in the  parameterization of Cugnon et al. \cite{Cugnon:1990xw}. Since we have also applied
the latter parameterization in the present calculations, this largely explains the
shift of our $(\Lambda+\Sigma^0)$ rapidity spectrum to smaller rapidities with respect to 
the measured spectrum.
Fig.~\ref{fig:dsig_dy_pbarTa_Lambda} demonstrates the sensitivity of our calculations
to the hyperon-nucleon cross sections.
\begin{figure}
\includegraphics[scale = 1.0]{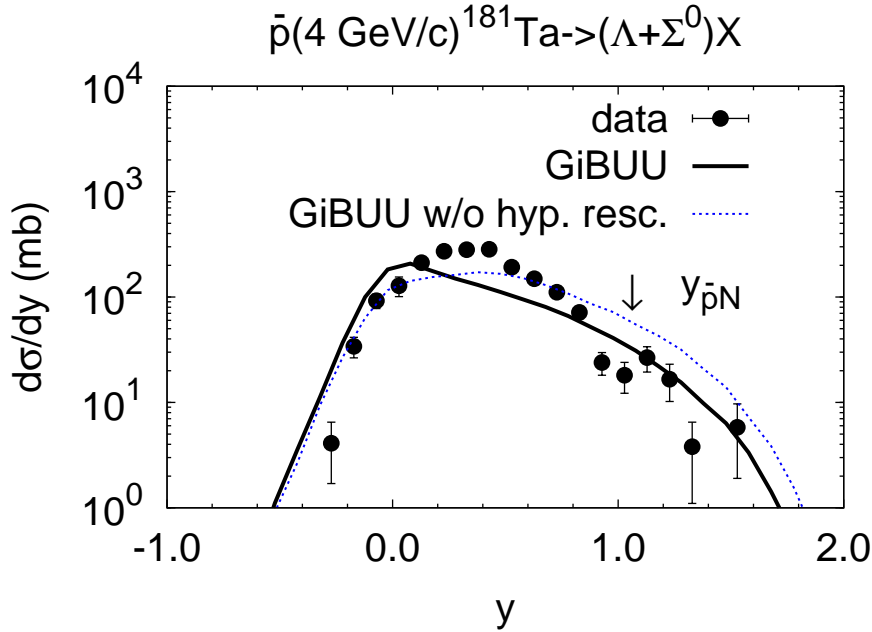}
\caption{\label{fig:dsig_dy_pbarTa_Lambda}
Rapidity distribution of $(\Lambda+\Sigma^0)$ for antiproton
collisions at 4 GeV/c with $^{181}$Ta calculated with (solid line) and
without (dotted line) taking into account hyperon-nucleon rescattering.
Experimental data are from \cite{Miyano:1988mq}.}
\end{figure}
The calculation without hyperon-nucleon rescattering produces the peak
position of the calculated $(\Lambda+\Sigma^0)$ rapidity spectrum close to the experimental one.
However, the spectrum at large forward rapidities is now overestimated. 
 
The stopping power of the nuclear medium with respect to the moving hyperon depends
not only on the integrated elastic hyperon-nucleon cross section, but also
on its angular dependence.
For simplicity, we have chosen the hyperon-nucleon cross sections to be isotropic 
in the c.m. frame. This also contributes to the somewhat too large deceleration
of the hyperons. 
On the other hand, the total yield of ($\Lambda+\Sigma^0$) hyperons can be enhanced
with increased charge exchange $\Sigma^- p \to \Lambda n$ and $\Sigma^+ n \to \Lambda p$ cross sections
which are rather poorly known from experiment.

Detailed information on the various production rates at 4 GeV/c is collected in Table~\ref{tab:pbarTa}.
\begin{table}[htb]
\caption{\label{tab:pbarTa} The same as Table \ref{tab:pbarNe},
but for reaction $\bar p$(4 GeV/c)+$^{181}$Ta.} 

\vspace{1cm}

\begin{tabular}{llllllll}
\hline
$\bar B B \to \Lambda X$     &3.6 (13.0)  &
$\bar B B \to \Sigma X$     &2.2 (0) &
$\bar B B \to \bar K X$      &109.9 (178.0)&
$\bar B B \to K X$         &110.1 (178.0)\\
$\pi B \to \Lambda K$        &6.6 (30.0)&
$\pi B \to \Sigma K$        &10.1 (36.0)& 
$\Sigma M \to \bar K X$      &1.1&
$\phi \to K \bar K$        &2.8 \\
$\eta B \to \Lambda K$      &1.2 (7.0)&
$\eta B \to \Sigma K$       &2.1 (4.0)&
$Y^* M \to \bar K X$        &3.8 &
$\pi B \to \Lambda K$       &6.6 (30.0)\\
$\rho B \to \Lambda K$      &2.5 &
$\rho B \to \Sigma K$       &6.0 &
$\phi \to K \bar K$        &2.8 &
$\eta B \to \Lambda K$      &1.2 (7.0)\\
$\omega B \to \Lambda K$     &2.0 (4.0)&
$\omega B \to \Sigma K$      &7.5 (8.0)&
$ M M \to K \bar K$        &32.4 &
$\rho B \to \Lambda K$      &2.5  \\
$\bar K B \to \Lambda X$     &20.1 (46.0)&
$\bar K B \to \Sigma X$      &13.4 (60.0)&
$K \bar K \to M M$         &-4.5 &
$\omega B \to \Lambda K$     &2.0 (4.0)\\
$\Sigma B \to \Lambda X$     &47.5 (76.0)&
$\Sigma B \to \Lambda X$     &-47.5 (-76.0)&
$\bar K B \to \Lambda X$     &-20.1 (-46.0)  &
$\pi B \to \Sigma K$        &10.1 (36.0)   \\
$Y^* B \to \Lambda X$       &11.4  &
$\Sigma B \to Y^* X$        &-1.3 &
$\bar K B \to \Sigma X$       &-13.4 (-60.0)  &
$\eta B \to \Sigma K$        &2.1 (4.0)\\
$\Lambda B \to \Sigma X$      &-12.5 (-26.0) &
$Y^* B \to \Sigma X$        &15.0  &
$\bar K B \to Y^*$          &-60.9  &
$\rho B \to \Sigma K$        &6.0 \\
$\Lambda B \to Y^* X$        &-2.5 &
$\Lambda B \to \Sigma X$      &12.5 (26.0) &
$Y^* \to \bar K N$         &22.4  &
$\omega B \to \Sigma K$      &7.5 (8.0)     \\
$\Lambda M \to Y^* X$        &-10.9  &
$\Sigma M \to Y^* X$         &-7.4 &
$\bar K B \to Y^* \pi$       &-25.0  &
$ M M \to K \bar K$         &32.4     \\
$Y^* \to \Lambda \pi(\eta)$    &33.4  &
$Y^* \to \Sigma \pi$        &27.2  &
                     &     &
$K \bar K \to M M$         &-4.5   \\
$B B \to \Lambda X$         &2.4 &
$\Sigma M \to \bar K X$      &-1.1 & 
                     &     &
$\bar Y B \to K X$         &3.8 (11.0) \\
                     & &
$B B \to \Sigma X$         &1.2 &
                     & &
$B B \to K X$            &3.9  \\
Sum $\Lambda$          &104.8 (150.0) &
Sum $\Sigma$           &39.9 (58.0) &
Sum $\bar K$           &48.5 (72.0) &
Sum $K$              &186.5 (278.0) \\
\hline
\end{tabular}
\end{table}
One can see from Table~\ref{tab:pbarTa} that with a relative contribution of $\sim 72\%$ the kaon absorption
reactions $\bar K B \to Y X$, $\bar K B \to Y^*$ and $\bar K B \to Y^* \pi$ are still dominating in the total
$Y$ and $Y^*$ production rate, which is somewhat smaller relative contribution than in the case of 
$\bar p$(607 MeV/c)$^{20}$Ne. This is expected, since at higher $\bar p$ beam momenta more strangeness production 
channels open.
The nonstrange meson-baryon collisions provide the second largest contribution of $\sim 23\%$ to the 
$Y$ and $Y^*$ production rate. The hyperon production in antibaryon-baryon collisions
(including direct channel) and in baryon-baryon collisions contributes only $\sim 3\%$ and $2\%$, respectively,
to the same rate. On the other hand, as before, in the case of $\bar p$(607 MeV/c)$^{20}$Ne, the antibaryon-baryon 
collisions dominate in the total $K$($\bar K$) production rate contributing $\sim 60\%$. It is interesting,
that the meson-meson reactions $M M \to K \bar K$ are rather important and contribute
$\sim 20\%$ to the $K$ and $\bar K$ production rates. This means that the $K \bar K$ pair
production processes in mesonic cloud created after $\bar p$ annihilation should also be taken 
into account on equal footing with other secondary production channels.
Note, however, that in GiBUU the two particles produced in the same two-body collision or resonance
decay event are allowed to collide only after at least one of them collided with another particle not involved 
in this event. This is done in order to avoid multiple chain reactions between the correlated products of
the same elementary event, which would otherwise lead to double counting in the production processes
and, moreover, violate the molecular chaos assumption. 

\begin{figure}
\includegraphics[scale = 0.80]{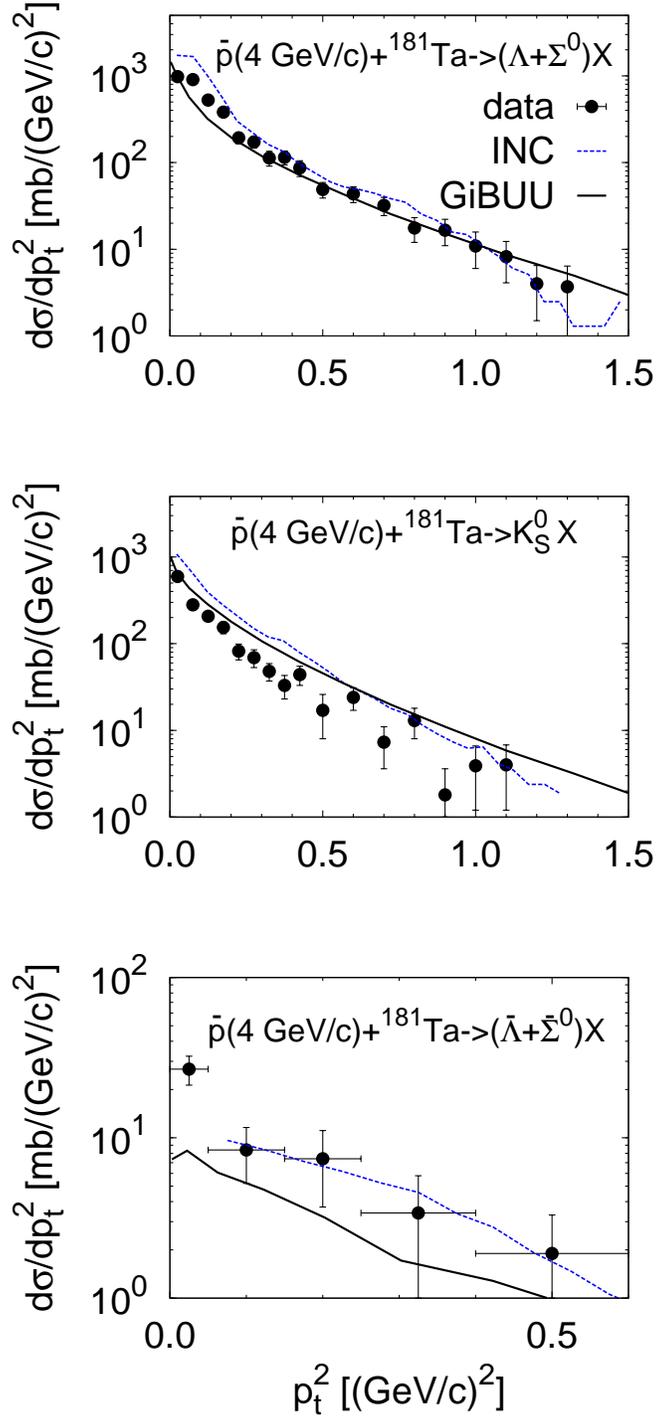}

\vspace*{0.5cm}

\caption{\label{fig:dsig_dpt2_pbarTa}
Inclusive transverse momentum spectra of ($\Lambda+\Sigma^0$), $K_S^0$ and ($\bar\Lambda + \bar\Sigma^0$)
from collisions $\bar p$+$^{181}$Ta at 4 GeV/c. The INC calculations from \cite{Cugnon:1990xw}
are also presented for comparison. 
The experimental data are taken from ref. \cite{Miyano:1988mq}.}
\end{figure}
Figure~\ref{fig:dsig_dpt2_pbarTa} shows the inclusive transverse momentum spectra of ($\Lambda+\Sigma^0$), $K_S^0$ 
and ($\bar \Lambda + \bar \Sigma^0$) from $\bar p$(4 GeV/c)+$^{181}$Ta collisions. 
The ($\Lambda+\Sigma^0$) production at low $p_t$ is clearly enhanced due to
large cross sections of the exothermic strangeness exchange processes $\bar K N \to Y \pi$ induced by slow antikaons. 
Moreover, the produced hyperons are decelerated by rescattering on nucleons. Our calculations reproduce the shapes
of the experimental $p_t^2$-spectra rather well. There is a remarkable agreement for
the absolute values of a ($\Lambda+\Sigma^0$) yield at high $p_t$. Again, the yield of $K_S^0$ is somewhat
overestimated, while the yield of antihyperons is underestimated by our calculations. 

In Table~\ref{tab:ssbar} we summarize the results of comparison of our calculations with INC
calculations and with experimental data at 0.6 and 4 GeV/c. 
Both models overestimate the strange quark production
and underestimate the ratio $\sigma_{\Lambda+\Sigma^0}/\sigma_{K_S^0}$.
\begin{table}[htb]
\caption{\label{tab:ssbar} The cross sections of the ($\Lambda+\Sigma^0$) hyperon,
$K_S^0$ meson and strange quark production (mb) in comparison with the INC results
from \cite{Cugnon:1990xw} and experimental data from \cite{Balestra:1987vy,Miyano:1988mq} 
for reactions  $\bar p$(607 MeV/c)+$^{20}$Ne and $\bar p$(4 GeV/c)+$^{181}$Ta. The ``experimental'' cross sections
of the $s$-quark production are, actually, evaluated in \cite{Cugnon:1990xw} based
on some simple relations to the really measured cross sections.}

\vspace{1cm}

\begin{tabular}{llllll}
\hline
system           &     &  $\sigma_{\Lambda+\Sigma^0}$  &  $\sigma_{K_S^0}$  &  $\sigma_{\Lambda+\Sigma^0}/\sigma_{K_S^0}$ & $\sigma_s$ \\
\hline
$\bar p$+$^{20}$Ne  &GiBUU&  18.7                         &  19.2              &  1.0                                        & 50.0 \\
                 &INC  &  19.6                         &  13.7              &  1.4                                        & 39.7 \\
                 &exp  &  $12.3 \pm 2.8$               &  $5.4 \pm 1.1$     &  $2.3 \pm 0.7$                              & 16.9 \\
$\bar p$+$^{181}$Ta &GiBUU&  154                          &  121               &  1.3                                        & 387 \\
                 &INC  &  275                          &  142               &  1.9                                        & 454 \\
                 &exp  &  $193 \pm 13$                 &  $82 \pm 6$        &  $2.4 \pm 0.3$                              & $277 \pm 21$ \\
\hline
\end{tabular}
\end{table}

\begin{figure}[ph]
\includegraphics[scale = 0.70]{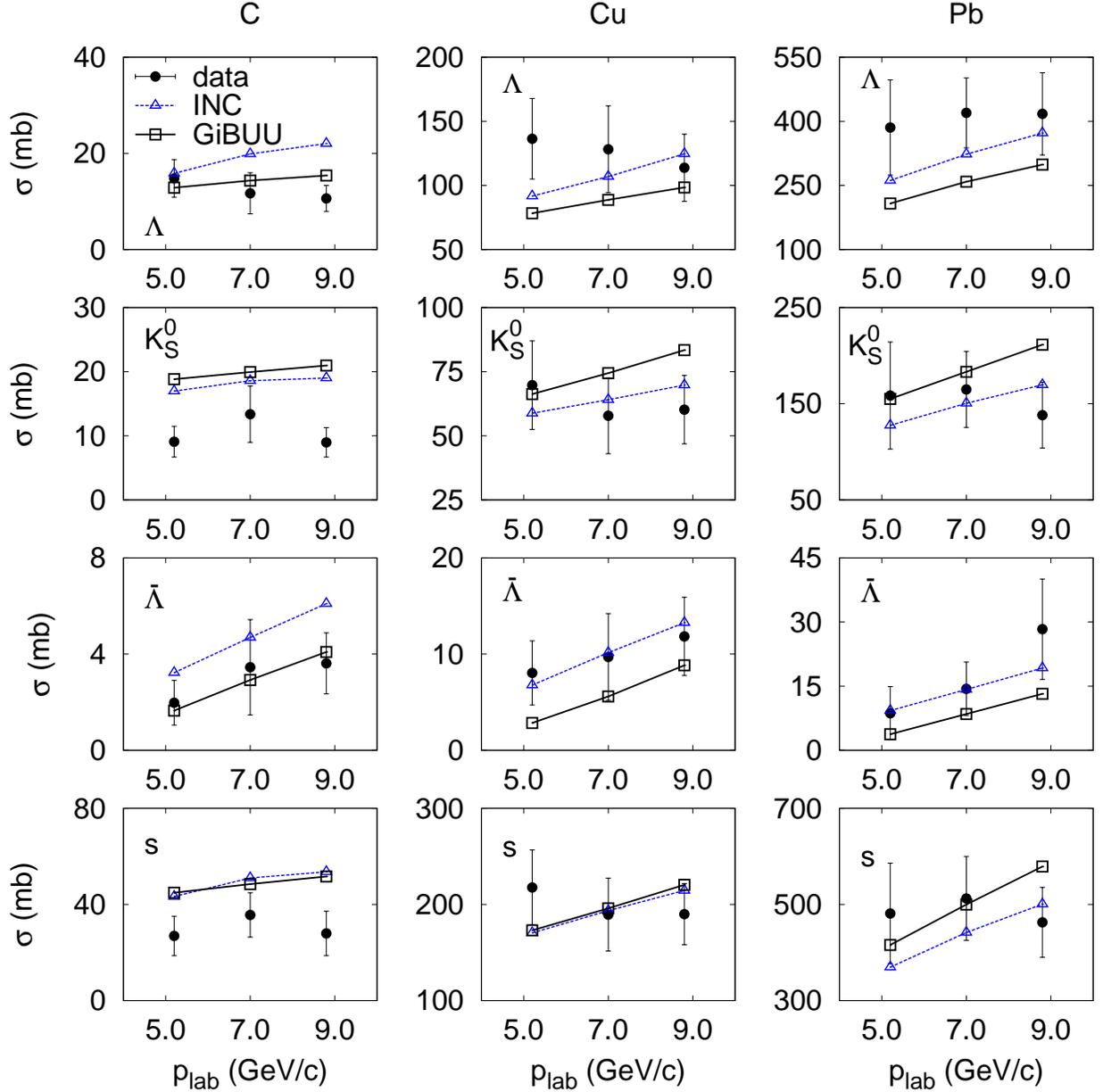}

\vspace*{0.5cm}

\caption{\label{fig:sig_str_vs_plab}
Inclusive cross section of $\Lambda$, $K_S^0$, $\bar\Lambda$ and 
strange quark (s) production for antiproton interactions 
at $p_{\text{lab}}=5.2$, 7.0 and 8.8 GeV/c with $^{12}$C, $^{64}$Cu and $^{208}$Pb targets.
The INC calculations and experimental data are from
ref. \cite{Ahmad:1997fv}.}
\end{figure}

\begin{figure}[ph]
\includegraphics[scale = 0.80]{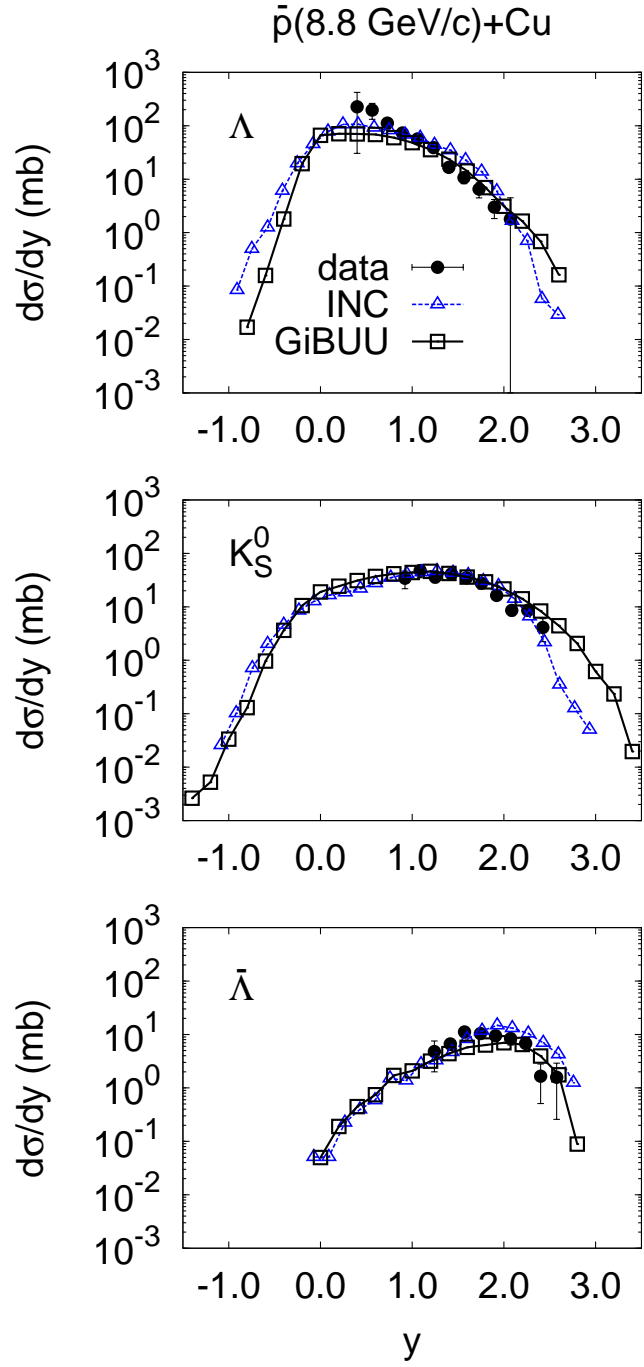}
\caption{\label{fig:dsigdy_pbarCu_9gevc}
Rapidity distributions of $\Lambda$, $K_S^0$ and $\bar\Lambda$
from $\bar p$ interactions with $^{64}$Cu at 8.8 GeV/c.
The INC calculations and experimental data are from
ref. \cite{Ahmad:1997fv}.}
\end{figure}

To our knowledge, the latest measurement of neutral strange particle production 
from high-energy antiproton interactions 
with nuclei was performed at BNL using the Multiparticle Spectrometer (MPS) facility
\cite{Ahmad:1997fv}. Fig.~\ref{fig:sig_str_vs_plab} shows the inclusive cross
sections for $\Lambda$, $K_S^0$, $\bar\Lambda$ and strange quark production in collisions of 
antiprotons at 5, 7 and 9 GeV/c with carbon, copper and lead targets
in comparison to the MPS data from \cite{Ahmad:1997fv}. Also the INC model
\cite{Strottman:1985jp,Gibbs:1990sf} results given in ref. \cite{Ahmad:1997fv}
are shown in Fig.~\ref{fig:sig_str_vs_plab}.
The strange quark (or $s\bar s$ pair) production cross section has been calculated
consistently with ref. \cite{Ahmad:1997fv}, i.e. according to the approximate formula
\begin{equation}
   \sigma_s=\frac{1}{2}(4\sigma_{K_S^0}+\sigma_\Lambda+\sigma_{\Sigma_0}
             +\sigma_{\bar\Lambda}+\sigma_{\bar\Sigma_0})~,
\end{equation}
where $\sigma_{K_S^0}=(\sigma_{K^0}+\sigma_{\bar K^0})/2$. 
There is a fair overall agreement of GiBUU calculations with data.
In particular, $\Lambda$ and $\bar\Lambda$ production on the carbon target is described very
well by GiBUU, while for the heavier targets we somewhat underpedict $\Lambda$ and $\bar\Lambda$ 
production. 
The $K_S^0$ production cross section on the carbon target is two times overpredicted by
GiBUU. On heavier targets, the agreement with experiment on $K_S^0$
production becomes better, but the slope of the beam momentum dependence of the 
$K_S^0$-production cross section seems to be overpredicted.
We note, that the data on the inclusive cross sections have been obtained
by integration over rapidity region with good acceptance and extrapolated
to the $4\pi$ solid angle using the INC calculations.
This is partly responsible for a better agreement of the INC results with this experiment.
Fig.~\ref{fig:dsigdy_pbarCu_9gevc} shows the rapidity distributions
of $\Lambda$, $K_S^0$ and
$\bar\Lambda$ from antiproton collisions
with copper target at 9 GeV/c. As one observes, the GiBUU calculations 
agree with the data quite well, except the underprediction of $\Lambda$
yield at $y \simeq 0.5$. We also observe a rather close agreement of GiBUU 
and INC results, which means that the selected
observables are, actually, not very sensitive to the model details,
once the elementary cross sections are adjusted to the experimental 
input.

\subsection{$S=-2$ hyperon production}
\begin{figure}
\includegraphics[scale = 0.80]{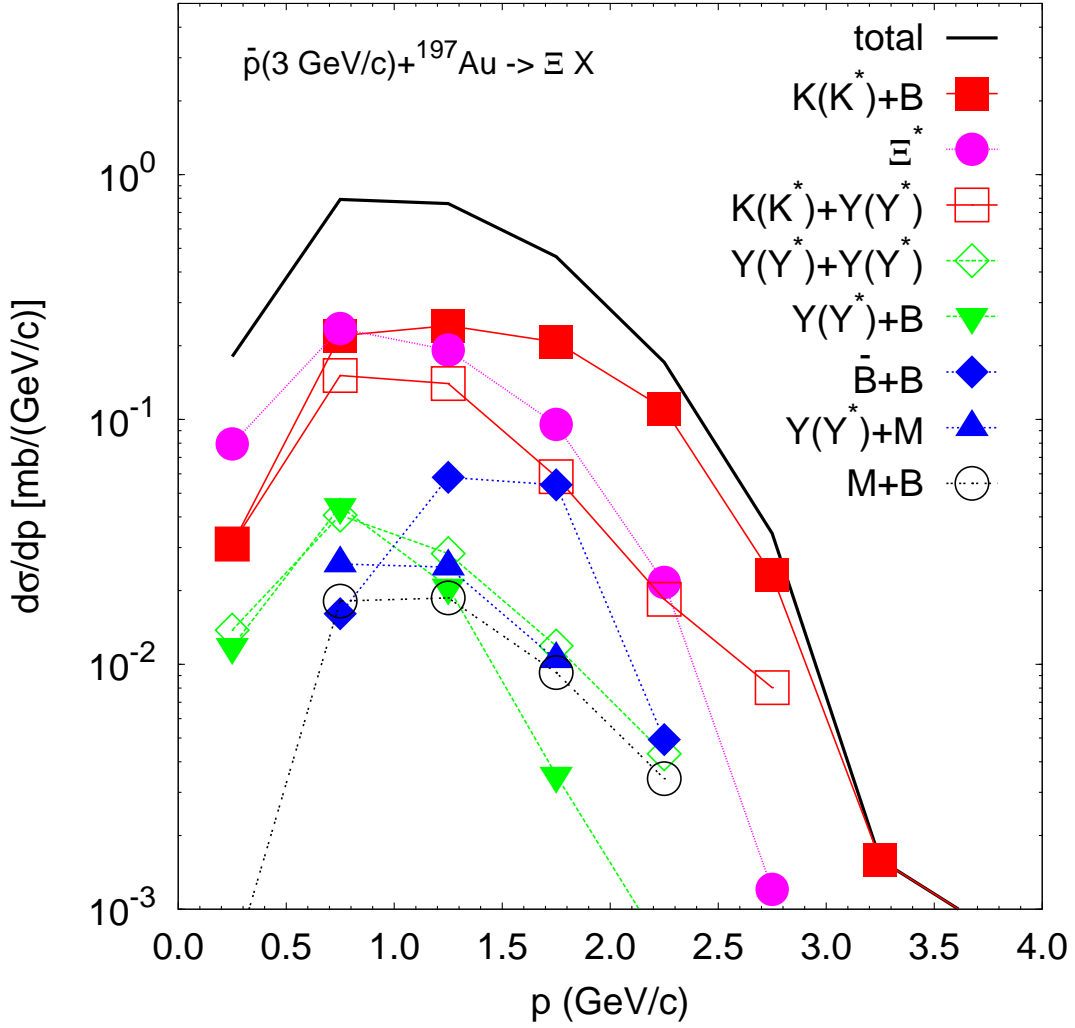}
\caption{\label{fig:dsig_dp_Xi_pbarAu_3gevc_decomp}
Inclusive momentum spectrum of $\Xi$ hyperons from 
$\bar p$(3 GeV/c)+$^{197}$Au collisions (the solid line). 
The partial contributions according to the parent particles
of a $\Xi$ hyperon are depicted by various lines with symbols.
In key notations, ``M'' denotes any nonstrange meson $\pi, \eta, \rho, \omega$,
``B'' --- any nonstrange baryon $N, \Delta, N^*$ etc.. The symbol ``$K(K^*)$''
denotes both kaon ($K^*$) and antikaon ($\bar K^*$).}
\end{figure}
Figure~\ref{fig:dsig_dp_Xi_pbarAu_3gevc_decomp} shows the inclusive momentum
spectrum of $\Xi$ hyperons from $\bar p$+$^{197}$Au collisions at 3 GeV/c
together with the partial contributions from various $\Xi$ production
channels. In performing this decomposition, we did not distinguish
kaons from antikaons. 
As one can see from Fig.~\ref{fig:dsig_dp_Xi_pbarAu_3gevc_decomp},
the (anti)kaon-baryon collisions deliver the main contribution $\sim 35\%$ to the $\Xi$ production,
mainly due to the double strangeness exchange channel $\bar K N \to K \Xi$. The decays
$\Xi^* \to \Xi \pi$ --- especially important at low transverse momenta of $\Xi$ --- make the 
second largest contribution $\sim 26\%$ to the $\Xi$ production. It is interesting, that
the (anti)kaon-hyperon collisions, which are collisions between the secondary particles,
contribute also quite appreciably, $\sim 17\%$. Other reaction channels are of relatively
minor importance for the inclusive $\Xi$ production. For example, the direct channel $\bar N N \to \bar\Xi \Xi$
contributes $\sim 6\%$ only; this channel is of primary importance for the
planned PANDA experiment on the double-$\Lambda$ hypernuclei production at FAIR
\cite{Pochodzalla:2005nf,Ferro:2007zz}.

\begin{figure}
\includegraphics[scale = 0.80]{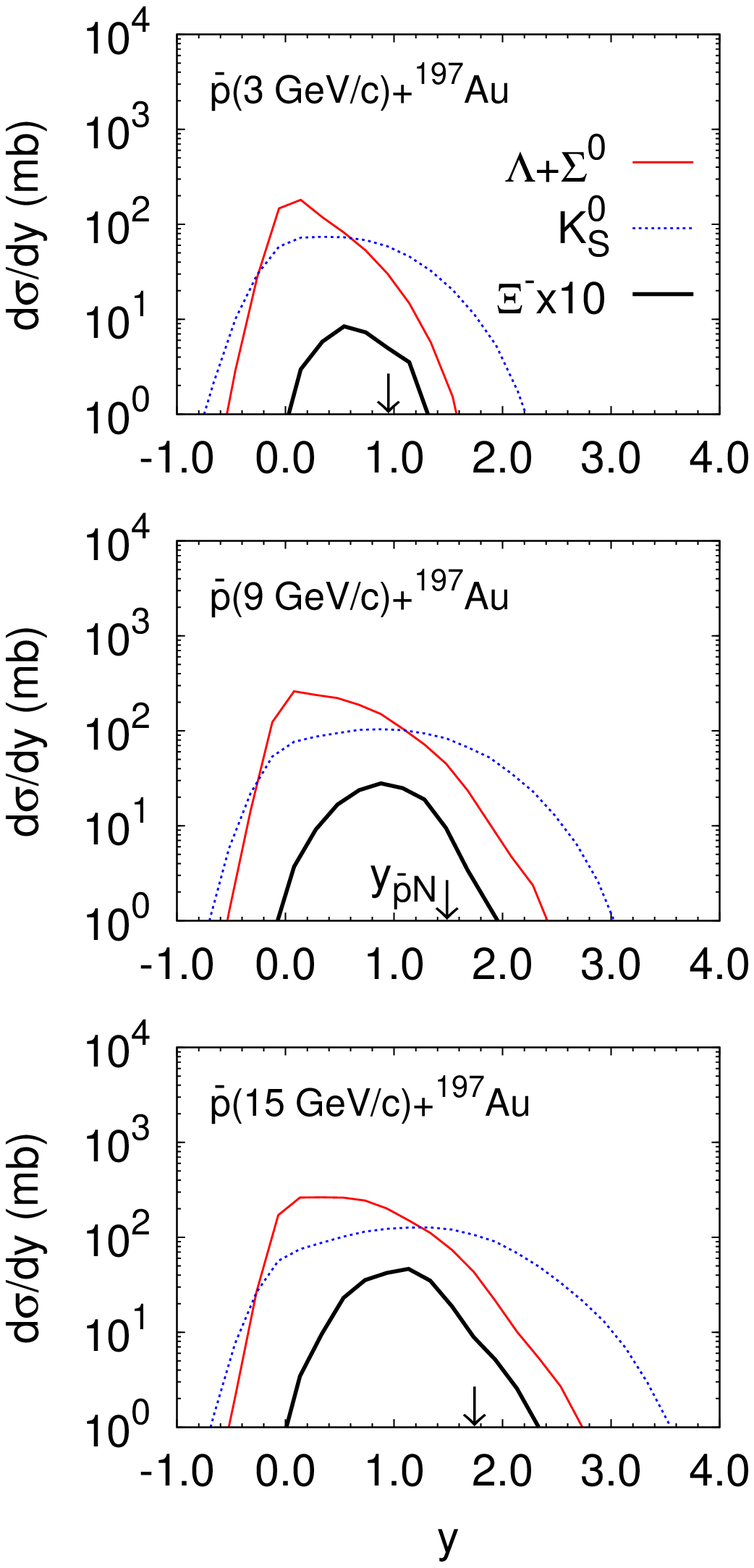}
\caption{\label{fig:dsig_dy_pbarAu}
Inclusive rapidity spectra of $\Xi^-$, ($\Lambda+\Sigma^0$) 
and $K^0_S$ from $\bar p$+$^{197}$Au collisions at 3, 9 and 15 GeV/c.
The spectra of $\Xi^-$ are multiplied by a factor of 10.
The $\bar p N$ c.m. rapidity $y_{\bar p N}=0.940,~1.479$ and $1.733$ for $p_{\rm lab}=3,~9$ and $15$ GeV/c,
respectively, is shown by vertical arrows.}
\end{figure}
Figure~\ref{fig:dsig_dy_pbarAu} shows the rapidity spectra of $\Xi^-$ 
hyperons together with the ($\Lambda+\Sigma^0$) and $K^0_S$ rapidity spectra from 
$\bar p$+$^{197}$Au collisions at 3, 9 and 15 GeV/c. The $\Xi^-$ spectra are about 
two orders of magnitude below those for the ($\Lambda+\Sigma^0$) and $K_S^0$ production. 
They are peaked at $y \simeq 0.5,~0.9$ and 1.2 for the beam momenta of 3, 9 and 15 GeV/c, 
respectively. However, the ($\Lambda+\Sigma^0$) spectra are always peaked near the target 
rapidity, $y = 0$, even at the largest beam momentum. This is because the hyperon 
production is dominated by the $\bar K N \to Y \pi$ processes with slow antikaons. 
Moreover, at 3 GeV/c, also the $K^0_S$ spectrum has a broad maximum at the target rapidity. 

The {\it experimental} fact that the ($\Lambda+\Sigma^0$) and $K^0_S$ rapidity spectra 
from $\bar p$+$^{181}$Ta collisions at 4 GeV/c are peaked
at nearly the same rapidities (c.f. Fig.~\ref{fig:dsig_dy_pbarTa_decomp}) has been
interpreted by Rafelski \cite{Rafelski:1988wn} as the manifestation of a common
production source for strange particles, i.e. of an annihilation fireball with
the baryon number $\simeq 10$. The large baryon number is due to absorbing nucleons
by the propagating fireball. Rafelski assumed that the fireball undergoes the transition
to the supercooled QGP state and then hadronizes. The rapidity spectra of
$\Xi$ hyperons would also be peaked at the fireball rapidity if the fireball
mechanism would dominate. In our model, a purely hadronic picture
emerges instead, where the $\Xi$ production is dominated by the double strangeness exchange
processes of $\bar K N \to K \Xi$ type. The latter are endothermic and require the momentum of
incoming antikaon in the rest frame of a nucleon placed above the threshold value $p_{\rm thr} \simeq 1.05$ 
GeV/c corresponding to the $\bar K N$ c.m. rapidity rapidity of 0.55. 
This makes the $\Xi$ rapidity spectra shift to higher rapidities.
In contrast, the $S=-1$ hyperon production is dominated by the exothermic strangeness
exchange $\bar K N \to \pi Y$. The cross section of this process grows with decreasing antikaon
momentum in the nucleon rest frame. This favors the isotropic production of slow hyperons in 
the target nucleus rest frame.

\begin{figure}
\includegraphics[scale = 0.80]{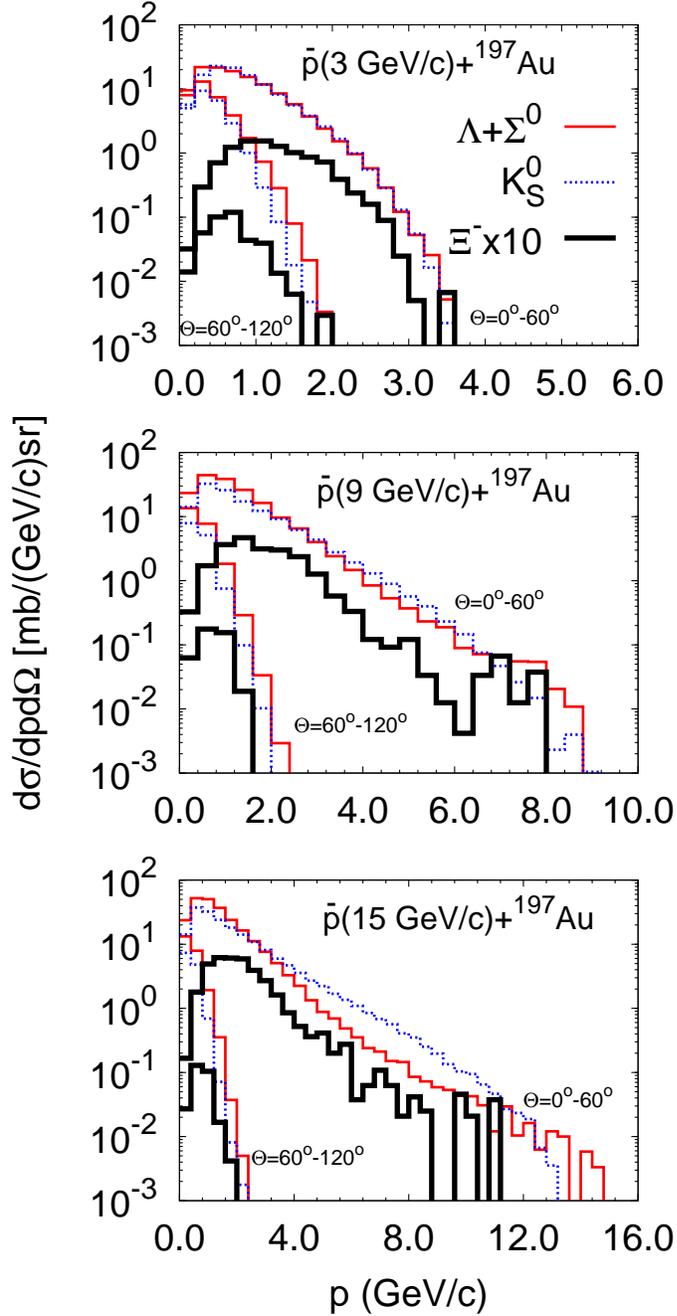}

\vspace*{0.5cm}

\caption{\label{fig:dsig_dp_pbarAu}
Inclusive laboratory momentum spectra of $\Xi^-$, ($\Lambda+\Sigma^0$) 
and $K^0_S$ from $\bar p$(3, 9, 15 GeV/c)+$^{197}$Au collisions 
in  the two ranges of the polar angle: $\Theta=0^o-60^o$   
and $\Theta=60^o-120^o$. The spectra of $\Xi^-$ are multiplied by a factor of 10.}
\end{figure}
Finally, Fig.~\ref{fig:dsig_dp_pbarAu} presents the laboratory momentum
spectra of the different strange particles in the forward ($\Theta=0^o-60^o$)
and transverse ($\Theta=60^o-120^o$) directions for $\bar p +^{197}$Au collisions 
at 3, 9 and 15 GeV/c. The spectra of ($\Lambda+\Sigma^0$) hyperons and $K^0_S$
are rather close to each other both in forward and transverse directions.
It is, moreover, interesting that the high-momentum slopes are similar for
all considered particles. However, the production of the low-momentum
$\Xi^-$ hyperons is suppressed due to the dominating $\Xi$ production
via the double strangeness exchange in $\bar K N \to K \Xi X$ processes.

\section{Summary and conclusions}
\label{summary}

To summarize, we have studied the strange particle production in $\bar p$-induced
reactions on nuclei by applying the GiBUU model. We have considered both at-rest
and in-flight reactions and confronted our results with experimental 
data \cite{Riedlberger:1989kn,Condo:1984ns,Balestra:1987vy,Miyano:1988mq,Ahmad:1997fv}
on neutral strange particle yields and spectra.  
We have also compared our
calculations with the earlier INC calculations of Cugnon-Deneye-Vandermeulen 
\cite{Cugnon:1990xw} and Gibbs-Kruk \cite{Gibbs:1990sf}. 
Finally, the model predictions for the $\Xi$ hyperon production are given.  

So far the main motivation for the experimental studies of strangeness production in
antiproton-nucleus reactions has been to find the signatures of 
QGP production, which can be complementary to similar studies in
high-energy heavy-ion collisions. Overall, our results are in a fair agreement 
with existing experimental data on $K^0_S$, 
($\Lambda+\Sigma^0$) and ($\bar\Lambda+\bar\Sigma^0$) production for $\bar p$ annihilation on nuclei
and with INC models.
There seems to be, indeed, no striking disagreements with data on inclusive 
$\Lambda$, $K_S^0$ and $\bar\Lambda$ production at high beam momenta, which could point to the 
exotic mechanisms, as it has been already concluded in refs. 
\cite{Gibbs:1990sf,Ahmad:1997fv}.

There are, however, some systematic deviations the origins of which remain to be better understood.
The GiBUU and INC models are based on the same hadronic cascade picture but differ
in several details like, e.g. the mean field potentials and
the treatment of $Y^*$ resonances. There is a clear tendency
to underestimate the $\Lambda/K^0_S$ ratio in our calculations.
This tendency is present also in the results of INC calculations, although 
somewhat less pronounced.
This problem might be caused by possible in-medium effects on the 
$\bar K N \to Y X$ cross section. Poorly known hyperon-nucleon cross sections 
also strongly influence the observed hyperon spectra. The latter cross
sections are also of more general interest, since they are related 
to the hyperon-nucleon interaction relevant in hypernuclear studies. 
To disentangle various possibilities, it would be very useful to measure 
the rapidity spectra of not only neutral strange particles, 
($\Lambda+\Sigma^0$) and $K_S^0$,  but also of the charged ones, $\Sigma^\pm$ and $K^\pm$.

The exotic mechanisms of strangeness production may, in-principle, manifest themselves 
more clearly in the $S=-2$ hyperon production. We have calculated the inclusive production 
of $\Xi$ hyperons from collisions $\bar p$+$^{197}$Au 
at 3-15 GeV/c. The $\Xi^-$ rapidity spectra
are strongly shifted to forward rapidities in the laboratory system,
since the $\Xi$ production via the dominating endothermic $\bar K N \to K \Xi$ 
channel requires a high-momentum initial antikaon. This is in a 
contrast to the strangeness production mechanism from a moving
annihilation fireball proposed by Rafelski \cite{Rafelski:1988wn}.
Thus a simultaneous measurement of the single- and double-strange hyperon
rapidity spectra may serve as a sensitive test for the hadronic and QGP scenarios
of strangeness production in $\bar p$ annihilation on nuclei.
The possibility for such measurements opens up in the planned 
PANDA experiment at FAIR \cite{PANDA}.

\begin{acknowledgments}
We thank H. Lenske, I.N. Mishustin, J. Pochodzalla, and I.A. Pshenichnov for their interest 
in our work and for stimulating discussions. We are especially grateful to\\
I.A. Pshenichnov for explanations on the application of his statistical annihilation code 
to the annihilation final states with strangeness.  
This work was financially supported by the Bundesministerium f\"ur Bildung und Forschung,
by the Helmholtz International Center for FAIR within the framework of the LOEWE program,
by DFG and by the Grant NSH-7235.2010.2 (Russia). 
The support by the Frankfurt Center for Scientific Computing is gratefully acknowledged.
\end{acknowledgments}

\appendix

\section{Strangeness production in $\bar N N$ annihilation at rest}
\label{strProd_atRest}

In this Appendix we list the probabilities for the various channels with strangeness for
$\bar p p$ and $\bar p n$ annihilation at rest (Tables~\ref{tab:brRat_ppbar} and \ref{tab:brRat_npbar},
respectively). We also provide references on source papers.
Footnotes explain how some of the probabilities have been obtained from experimental data. 
Many of the partial probabilities for the $\bar p p$ annihilation at rest
can be also found in the data compilation \cite{Bluem88}.

\newpage

\begin{table}[htb]
\caption{\label{tab:brRat_ppbar} Probabilities of the different channels with strange mesons 
for $\bar p p$ annihilation at rest (in \%).}
\begin{ruledtabular}
\begin{tabular}{llllll}
  Channel             &  Prob.  &  Ref. &  Channel              &  Prob.  &  Ref.  \\ 
\hline
  $K^+ K^-$~$^{a)}$      &  0.091        & \cite{Amsler:1992ku,Doser:1988fw}  &      $K^{*+} K^- \pi^0$          & 0.070        &  \cite{PhysRev.145.1095} \\
  $K^0 \bar K^0$~$^{a)}$   &  0.091        & \cite{Amsler:1992ku,Doser:1988fw}  &      $K^{*-} K^+ \pi^0$          & 0.070        &  \cite{PhysRev.145.1095} \\
  $K^+ K^{*-}$          &  0.071        & \cite{Soulliere:1987in}            &      $\bar K^{*0} K^0 \pi^0$~$^{b)}$ & 0.070        &  \\
  $K^- K^{*+}$          &  0.071        & \cite{Soulliere:1987in}            &      $K^{*0} \bar K^0 \pi^0$~$^{b)}$ & 0.070        &  \\
  $K^0 \bar K^{*0}$       &  0.060        & \cite{Barash:1965zz}               &      $K^{*0} K^- \pi^+$          & 0.085        &  \cite{PhysRev.145.1095} \\
  $\bar K^0 K^{*0}$       &  0.060        & \cite{Barash:1965zz}               &      $\bar K^{*0} K^+ \pi^-$       & 0.085        &  \cite{PhysRev.145.1095} \\
  $K^{*+} K^{*-}$        &  0.225        & \cite{PhysRev.145.1095}            &      $K^0 \bar K^0 \pi^0 \pi^0$~$^{6)}$  & 0.035     &  \cite{Abele:1997vu,PhysRev.145.1095} \\
  $K^{*0} \bar K^{*0}$     &  0.225        & \cite{PhysRev.145.1095}            &      $K^+ K^- \pi^0 \pi^0$~$^{b)}$     & 0.035     &  \\
  $K^0 \bar K^0 \pi^0$~$^{1)}$  &  0.146        & \cite{Barash:1965zz}            &      $K^0 \bar K^0 \pi^+ \pi^- \pi^0$~$^{7)}$ & 0.068 & \cite{Barash:1900zz,Reifenroether:1991ik} \\ 
  $K^+ K^- \pi^0$~$^{b)}$     &  0.146        &                                 &      $K^+ K^- \pi^+ \pi^- \pi^0$~$^{b)}$     & 0.068      & \\ 
  $K^0 K^- \pi^+$~$^{2)}$     &  0.142        & \cite{Barash:1965zz}            &      $K^0 K^- \pi^- \pi^+ \pi^+$~$^{8)}$     & 0.059      & \cite{Barash:1900zz} \\
  $\bar K^0 K^+ \pi^-$~$^{2)}$  &  0.142        & \cite{Barash:1965zz}            &      $\bar K^0 K^+ \pi^+ \pi^- \pi^-$~$^{8)}$  & 0.059      & \cite{Barash:1900zz} \\
  $K^0 \bar K^0 \eta$~$^{3)}$   & 0.050        &  \cite{Barash:1900zz}            &      $K^0 K^- \pi^+ \pi^0 \pi^0$~$^{b)}$     & 0.042      & \\ 
  $K^+ K^- \eta$~$^{b)}$      & 0.050        &                                  &      $\bar K^0 K^+ \pi^- \pi^0 \pi^0$~$^{b)}$  & 0.042      & \\ 
  $K^0 \bar K^0 \omega$~$^{3,a)}$  & 0.232        &  \cite{Bizzari:1971ta}          &      $K^+ K^- \pi^0 \pi^0 \pi^0$~$^{b)}$     & 0.012      & \\ 
  $K^+ K^- \omega$~$^{a)}$     & 0.232        &  \cite{Bizzari:1971ta}           &      $\bar K^0 K^0 \pi^0 \pi^0 \pi^0$~$^{b)}$  & 0.012      & \\ 
  $K^0 \bar K^0 \rho^0$~$^{4,a)}$  & 0.202         &  \cite{Amsler:1986cr,Heel:1986aq,PhysRev.145.1095,Reifenroether:1991ik} &
                         $\phi \pi^+ \pi^-$             & 0.054        &  \cite{Reifenroether:1991ik} \\                       
  $K^+ K^-  \rho^0$~$^{4,a)}$    & 0.202        &  \cite{Amsler:1986cr,Heel:1986aq,PhysRev.145.1095,Reifenroether:1991ik} &
                         $\phi \pi^0 \pi^0$~$^{b)}$       & 0.011     &     \\                         
  $K^0 K^-  \rho^+$~$^{5)}$    & 0.234        &  \cite{PhysRev.145.1095} &
                         $\phi \rho^0$                & 0.034        &  \cite{Reifenroether:1991ik} \\                         
  $\bar K^0 K^+  \rho^-$~$^{5)}$ & 0.234         &  \cite{PhysRev.145.1095} &
                         $\phi \pi^0$                & 0.019        &  \cite{Reifenroether:1991ik} \\                         
  $K^{*+} \bar K^0 \pi^-$       & 0.230        &  \cite{PhysRev.145.1095} &
                         $\phi \eta$                 & 0.004          & \cite{Reifenroether:1991ik} \\                         
  $K^{*-} K^0 \pi^+$          & 0.230        &  \cite{PhysRev.145.1095} &
                         $\phi \omega$                & 0.030          & \cite{Reifenroether:1991ik} \\
\end{tabular}
\end{ruledtabular}
\end{table}

$^{1)}$Nonresonance contribution obtained as
$B_{K^0 \bar K^0 \pi^0}^{\rm nonres}=B_{K_S K_S \pi_0}+B_{K_L K_L \pi_0}=1.46\cdot10^{-3}$.

$^{2)}$Nonresonance contributions obtained as
\begin{eqnarray*}
 &&B_{K^0 K^- \pi^+}^{\rm nonres}=B_{\bar K^0 K^+ \pi^-}^{\rm nonres} 
 =0.5\cdot(B_{K^0 K^\pm \pi^\mp}-B_{K^0 K^{*0}, K^{*0} \to K^\pm \pi^\mp}
-B_{K^\pm K^{*\mp}, K^{*\mp} \to K^0 \pi^\mp}) \\
 && = 0.5\cdot(4.25-0.85-0.57)\cdot10^{-3}=1.42\cdot10^{-3}~.
\end{eqnarray*}  

$^{3)}$Obtained as $B_{K^0 \bar K^0 M}=B_{K_S K_S M}+B_{K_L K_L M} $, where $M=\eta,~\omega$. 

$^{4)}$Obtained as
\begin{eqnarray*}
   && B_{K^0 \bar K^0 \rho^0} = B_{K^0 \bar K^0 \pi^+ \pi^-}
     -B_{K^{*\pm} K^0 \pi^\mp, K^{*\pm} \to K^0 \pi^\pm}
     -B_{K^{*+} K^{*-}, K^{*+} \to K^0 \pi^+, K^{*-} \to \bar K^0 \pi^-} \\
   && -0.34 \cdot (B_{\phi \pi^+ \pi^-} + B_{\phi \rho^0})
     = (7.45 - 3.63 - 0.57 - 0.34\cdot(0.54+0.34))\cdot10^{-3}
      = 2.96\cdot10^{-3}~, \\
   && B_{K^+ K^- \rho^0} = B_{K^+ K^- \pi^+ \pi^-}
     -B_{K^{*0} \bar K^{*0}, K^{*0} \to K^+ \pi^-, \bar K^{*0} \to K^- \pi^+}
     -\frac{2}{3}B_{K^{*0} K^\pm \pi^\mp} - 0.49 \cdot (B_{\phi \pi^+ \pi^-}
      + B_{\phi \rho^0}) \\
   && = (3.6-1.0-\frac{2}{3}\cdot1.7-0.49\cdot(0.54+0.34))\cdot10^{-3}
      =1.07\cdot10^{-3}~.
\end{eqnarray*}

$^{5)}$Obtained as
\begin{eqnarray*}
   && B_{K^0 K^\pm \rho^\mp} = B_{K^0 K^\pm \pi^\mp \pi^0}
        - B_{K^{*0} K^\pm \pi^\mp, K^{*0} \to K^0 \pi^0} 
        - B_{K^{*\mp} K^\pm \pi^0, K^{*\mp} \to K^0 \pi^\mp}
        - B_{K^{*\pm} K^0 \pi^\mp, K^{*\pm} \to K^\pm \pi^0} \\
   &&   - B_{K^{*+} K^{*-}, K^{*+} \to K^0 \pi^+ (K^+ \pi^0),
             K^{*-} \to K^- \pi^0 (\bar K^0 \pi^-)}
        - B_{K^{*0} \bar K^{*0}, K^{*0} \to K^0 \pi^0 (K^+ \pi^-),
             \bar K^{*0} \to K^- \pi^+ (\bar K^0 \pi^0)} \\
   &&   - B_{\bar K^{*0} K^0 \pi^0, \bar K^{*0} \to K^- \pi^+}
        - B_{K^{*0} \bar K^0 \pi^0, K^{*0} \to K^+ \pi^-} \\
   && = (10.38-0.57-0.97-1.04-0.77-1.43-\frac{2}{3}\cdot1.4)\cdot10^{-3}
      =4.67\cdot10^{-3}~.
\end{eqnarray*}

$^{6)}$Nonresonance contribution obtained  as
\begin{eqnarray*}
   && B_{K^0 \bar K^0 \pi^0 \pi^0}^{\rm nonres}=B_{K^0 \bar K^0 \pi^0 \pi^0}
         -\frac{1}{3}(B_{K^{*0} \bar K^0 \pi^0}+B_{\bar K^{*0} K^0 \pi^0})
  - \frac{1}{9} B_{K^{*0} \bar K^{*0}}-0.34B_{\phi \pi^0 \pi^0} \\
   && = (1.1-\frac{1}{3}\cdot1.4-\frac{1}{9}\cdot2.25-0.34\cdot0.11)\cdot10^{-3}
=0.35\cdot10^{-3}~,
\end{eqnarray*}
assuming that $B_{K^0 \bar K^0 \pi^0 \pi^0}=B_{K_L K_S \pi^0 \pi^0}$ according to \cite{Abele:1997vu}.

$^{7)}$Nonresonance contribution obtained  as
\begin{eqnarray*}
   && B_{K^0 \bar K^0 \pi^+ \pi^- \pi^0}^{\rm nonres}
   = B_{K^0 \bar K^0 \pi^+ \pi^- \pi^0} - 0.28 \cdot B_{K^0 \bar K^0 \eta} 
     - 0.89 \cdot B_{K^0 \bar K^0 \omega} - 0.34 \cdot (0.28 \cdot B_{\phi \eta}
     + 0.89 \cdot B_{\phi \omega}) \\
   && = (2.98 - 0.28 \cdot 0.50 - 0.89 \cdot 2.32 -0.34 \cdot (0.28 \cdot 0.036
      + 0.89 \cdot 0.30))\cdot10^{-3} = 0.68\cdot10^{-3}~.
\end{eqnarray*}
   
$^{8)}$Obtained as $B_{\bar K^0 K^+ \pi^+ \pi^- \pi^-}=B_{K^0 K^- \pi^- \pi^+ \pi^+}=B_{K_S K^\pm \pi^\pm \pi^\mp \pi^\mp}$.  

$^{a)}$Averaged with respect to the charged and neutral $K$ or $K^*$ production
channels.

$^{b)}$Obtained by isospin relations from the branching ratio of another channel.

\newpage

\begin{table}[htb]
\caption{\label{tab:brRat_npbar} Probabilities of the different channels with strange mesons 
for $\bar p n$ annihilation at rest (in \%).}
\begin{ruledtabular}
\begin{tabular}{llllll}
  Channel             &  Prob.  &  Ref. &    Channel             &  Prob.  &  Ref. \\ 
\hline
  $K^0 K^-$              &  0.147        & \cite{Bettini:1969ki} &       $\bar{K^{*0}} K^0 \pi^-$~$^{7)}$ & 0.130      & \cite{Bettini:1969ki} \\ 
  $K^0 K^{*-}$~$^{1)}$      &  0.067        & \cite{Bettini:1969sy} &       $K^{*+} K^- \pi^-$~$^{8)}$     & 0.154      & \cite{Bettini:1969ki} \\ 
  $K^- K^{*0}$~$^{1)}$      &  0.067        & \cite{Bettini:1969sy} &       $K^{*-} K^+ \pi^-$~$^{8)}$     & 0.154      & \cite{Bettini:1969ki} \\ 
  $K^{*-} K^{*0}$~$^{2)}$    &  0.184        & \cite{Bettini:1969ki} &       $K^0 K^- \pi^0 \pi^0$~$^{a)}$   & 0.043      &    \\
  $K^0 K^- \pi^0$~$^{3)}$    &  0.316        & \cite{Bettini:1969sy} &       $\bar{K^0} K^+ \pi^- \pi^- \pi^0$    & 0.016      & \cite{Bettini:1969ki} \\ 
  $K^0 \bar{K^0} \pi^-$~$^{4)}$ & 0.432        & \cite{Bettini:1969sy} &       $\bar{K^0} K^0 \pi^- \pi^- \pi^+$      & 0.075      & \cite{Bettini:1969ki} \\
  $K^+ K^- \pi^-$~$^{5)}$    & 0.513        & \cite{Bettini:1969sy}  &       $K^- K^+ \pi^- \pi^- \pi^+$~$^{a)}$    & 0.075      &                       \\
  $K^0 K^- \eta$~$^{a)}$     & 0.070       &                        &        $\bar{K^0} K^0 \pi^0 \pi^0 \pi^-$~$^{a)}$ & 0.052     &                        \\
  $K^0 K^- \omega$          & 0.350       & \cite{Bettini:1969ki}  &        $K^+ K^- \pi^0 \pi^0 \pi^-$~$^{a)}$    & 0.052     &                        \\
  $K^0 K^- \rho^0$~$^{6)}$     & 0.150     & \cite{Bettini:1969ki}   &        $K^0 K^- \pi^+ \pi^- \pi^0$~$^{9)}$  & 0.025      & \cite{Bettini:1969ki} \\
  $K^0 \bar{K^0} \rho^-$~$^{7)}$ & 0.770       & \cite{Bettini:1969ki} &        $K^0 K^- \pi^0 \pi^0 \pi^0$~$^{a)}$  & 0.015      &                       \\
  $K^+ K^- \rho^-$~$^{a)}$    & 0.770       &                        &        $\phi \pi^- \pi^0$~$^{a)}$     & 0.065        &     \\
  $K^{*-} K^0 \pi^0$~$^{7)}$     & 0.245      & \cite{Bettini:1969ki} &        $\phi \rho^-$~$^{a)}$        & 0.068        &     \\
  $K^{*0} K^- \pi^0$~$^{a)}$     & 0.245      &                       &        $\phi \pi^-$               & 0.088        & \cite{Bettini:1969sy} \\
  $K^{*0} \bar{K^0} \pi^-$~$^{7)}$ & 0.130      & \cite{Bettini:1969ki} &                               &              &  \\ 
\end{tabular}
\end{ruledtabular}
\end{table}

$^{1)}$Obtained by averaging the transition rates into $K^* \bar K + \bar{K^*} K$
from $^1S_0~I=1$ and $^3S_1~I=1$ states as
\begin{eqnarray*}
   && B_{K^* \bar K + \bar{K^*} K} = 0.5\cdot(2.0+24.6)\cdot10^{-4}
     =13.3\cdot10^{-4}~, \\
   && B_{K^0 K^{*-}}=B_{K^- K^{*0}}=0.5 \cdot B_{K^* \bar K + \bar{K^*} K}
     =6.7\cdot10^{-4}~.
\end{eqnarray*}

$^{2)}$Reconstructed from the partial contribution into the 
$\bar{K^0} K^+ \pi^- \pi^-$ channel.

$^{3)}$Nonresonance contribution calculated as
\begin{eqnarray*}
   && B_{K^0 K^- \pi^0}^{\rm nonres}=B_{K^0 K^- \pi^0}
      -B_{K^0 K^{*-}, K^{*-} \to K^- \pi^0}
      -B_{K^- K^{*0}, K^{*0} \to K^0 \pi^0} \\
   && = B_{K^0 K^- \pi^0} - \frac{1}{3} \cdot B_{K^0 K^{*-}+K^- K^{*0}}
      = (36.0 - \frac{1}{3} \cdot 13.3)\cdot10^{-4}=31.6\cdot10^{-4}~.
\end{eqnarray*}

$^{4)}$Nonresonance contribution evaluated as
\begin{eqnarray*}
   && B_{K^0 \bar{K^0} \pi^-}^{\rm nonres}=B_{K^0 \bar{K^0} \pi^-}
     -B_{K^0 K^{*-}, K^{*-} \to \bar{K^0} \pi^-} 
     -B_{\phi \pi^-, \phi \to K_L^0 K_S^0} \\
   && =(50.6-\frac{2}{3}\cdot6.7-0.34\cdot8.8)\cdot10^{-4}
      =43.2\cdot10^{-4}~.
\end{eqnarray*}

$^{5)}$Nonresonance contribution evaluated as
\begin{eqnarray*}
   && B_{K^+ K^- \pi^-}^{\rm nonres}=B_{K^+ K^- \pi^-}
     -B_{K^- K^{*0}, K^{*0} \to K^+ \pi^-}
     -B_{\phi \pi^-, \phi \to K^+ K^-} \\
   && =(60.0-\frac{2}{3}\cdot6.7-0.49\cdot8.8)\cdot10^{-4}
      =51.3\cdot10^{-4}~. 
\end{eqnarray*}

$^{6)}$Reconstructed from the partial contribution to
the $K^0 K^- \pi^- \pi^+$ channel. 

$^{7)}$Reconstructed from the partial contribution to
the $K^0_S K^0_S \pi^- \pi^0$ channel.

$^{8)}$Reconstructed from the partial contribution to
the $K^0 K^- \pi^+ \pi^-$ and $\bar{K^0} K^+ \pi^- \pi^-$ channels. 

$^{9)}$Nonresonance contribution evaluated as
\begin{eqnarray*}
   && B_{K^0 K^- \pi^+ \pi^- \pi^0}^{\rm nonres}=B_{K^0 K^- \pi^+ \pi^- \pi^0}
     -B_{K^0 K^- \omega, \omega \to \pi^+ \pi^- \pi^0} \\
   && =(33.6-0.89\cdot35.0)\cdot10^{-4}
      =2.5\cdot10^{-4}~.
\end{eqnarray*}

$^{a)}$Obtained by isospin relations from the branching ratio of
another channel. 

\newpage

\section{Elementary cross sections}
\label{elem}

In this Appendix we describe the cross sections of the hyperon production in 
$\bar N N$ collisions. Other partial cross sections of the $\bar N N$ collisions 
are described in Appendix B.2 of ref. \cite{Buss:2011mx}.

The $\bar p p \to \bar\Lambda \Lambda$ and $\bar p p \to \bar\Lambda \Sigma^0+(\mbox{c.c.})$ cross sections (in mb) are parameterized 
as a function of the invariant energy $\sqrt{s}$ (in GeV) by the following expressions:
\begin{equation}
  \sigma_{\bar p p \to \bar\Lambda \Lambda}=
  \cases{
         0, & $\sqrt{s} < 2.232$ \cr 
         0.0357(\sqrt{s}-2.232)^{0.5}+9.005(\sqrt{s}-2.232)^{1.5}, & $2.232 < \sqrt{s} < 2.252$ \cr  
         \mbox{max}(0.01,0.726(s/4.982-1)^{0.774} (4.982/s)^{3.350}), & $2.252 < \sqrt{s}$
  }
  \label{sig_pbarp_to_LambdaBarLambda}
\end{equation}
\begin{equation}
  \sigma_{\bar p p \to \bar\Lambda \Sigma^0 + \mbox{c.c.}} =
  \cases{
         0~, &~~~$\sqrt{s} < 2.308$ \cr
         0.184(s/5.327-1)^{0.437} (5.327/s)^{1.850}~, &~~~$2.308 < \sqrt{s}$~.
  }
  \label{sig_pbarp_to_LambdaBarSigma}
\end{equation}
The functional form in (\ref{sig_pbarp_to_LambdaBarLambda}) at lower energies
is adopted from ref. \cite{Barnes:1989je}, however, with slightly different 
numerical parameters. The parameterization used in (\ref{sig_pbarp_to_LambdaBarLambda}) at higher energies and
in (\ref{sig_pbarp_to_LambdaBarSigma}) is taken from ref. \cite{Tsushima:1998jz}.
The numerical parameters in (\ref{sig_pbarp_to_LambdaBarLambda})
and (\ref{sig_pbarp_to_LambdaBarSigma}) are obtained by the fit to the world data on the cross sections 
$\bar p p \to \bar\Lambda \Lambda$ and $\bar p p \to \bar\Lambda \Sigma^0+\mbox{c.c.}$ from \cite{Baldini:1988ti}.

For the $\bar p p \to \bar\Xi^+ \Xi^-$ cross section only the value of $2$ $\mu$b at 3 GeV/c is known experimentally
\cite{Baldini:1988ti}. Thus, we simply assume the constant cross section above threshold:
\begin{equation}
  \sigma_{\bar p p \to \bar\Xi^+ \Xi^-} = \sigma_{\bar p p \to \bar\Xi^0 \Xi^0} =
  \cases{
     0~\mu\mbox{b}, & $\sqrt{s} < 2.630$ GeV \cr
     2~\mu\mbox{b}, & $2.630$ GeV $< \sqrt{s}$~.
  }
\end{equation}

The cross sections involving $n,~\bar n,~\Delta$ and $\bar\Delta$ in the initial state are
obtained from the isospin relations as
\begin{eqnarray}
   \sigma_{\bar n n \to \bar\Lambda \Lambda} &=& \sigma_{\bar p p \to \bar\Lambda \Lambda}~,\\
   \sigma_{\bar B B \to \bar\Lambda \Sigma + \mbox{c.c.}} &=&
       2 \sigma_0
       \clebsch{I^{\bar B}}{I_z^{\bar B}}{I^B}{I_z^B}{1}{I_z^\Sigma}~, \label{sig_BbarB_to_LbarS} \\
   \sigma_{\bar B B \to \bar\Xi \Xi} &=& 
       \sigma_1
       \sum_{I=0,1}\,\clebsch{I^{\bar B}}{I_z^{\bar B}}{I^B}{I_z^B}{I}{(I_z^{\bar B}+I_z^B)}
                     \clebsch{\frac{1}{2}}{I_z^{\bar\Xi}}{\frac{1}{2}}{I_z^\Xi}{I}{(I_z^{\bar B}+I_z^B)}
                                                                       \label{sig_BbarB_to_XibarXi}
\end{eqnarray}
where $\sigma_0 = \sigma_{\bar p p \to \bar\Lambda \Sigma^0 + \mbox{c.c.}}$, 
$\sigma_1 = 2 \sigma_{\bar p p \to \bar\Xi^+ \Xi^-}$,
$B$ stands for $N$ or $\Delta$,
and only the combinations $\bar N N$, $\bar\Delta N$ and $\bar N \Delta$     
of incoming particles are taken into account. Equations (\ref{sig_BbarB_to_LbarS}) and 
(\ref{sig_BbarB_to_XibarXi}) can be simplified, which gives the following:
\begin{eqnarray}
   \sigma_{\bar n n \to \bar\Lambda \Sigma^0 + \mbox{c.c.}} &=& \sigma_0 \label{sig_nbarn_to_LbarSig0}  ~,\\
   \sigma_{\bar p n \to \bar\Lambda \Sigma^- + \Lambda \bar\Sigma^-} &=& 2\sigma_0~,\\
   \sigma_{\bar\Delta^- p \to \bar\Lambda \Sigma^0 + \mbox{c.c.}} 
   = \sigma_{\bar\Delta^0 n \to \bar\Lambda \Sigma^0 + \mbox{c.c.}} &=& \sigma_0 \label{sig_D1}~,\\
   \sigma_{\bar\Delta^- n \to \bar\Lambda \Sigma^- + \Lambda \bar\Sigma^-} 
   = \sigma_{\bar\Delta^0 p \to \bar\Lambda \Sigma^+ + \Lambda \bar\Sigma^+} &=& \frac{1}{2} \sigma_0  \label{sig_D2}~,\\
   \sigma_{\bar\Delta^{--} p \to \bar\Lambda \Sigma^- + \Lambda \bar\Sigma^-}
   = \sigma_{\bar\Delta^+ n \to \Lambda\bar\Sigma^+ + \bar\Lambda \Sigma^+} &=& \frac{3}{2} \sigma_0 \label{sig_D3}~\\
   \sigma_{\bar n n \to \Xi^-\bar\Xi^+ + \Xi^0\bar\Xi^0}
   = \sigma_{\bar p n \to \Xi^-\bar\Xi^0} &=& \sigma_1 ~,\\
   \sigma_{\bar\Delta^- p \to \Xi^-\bar\Xi^+ + \Xi^0\bar\Xi^0}
   = \sigma_{\bar\Delta^0 n \to \Xi^-\bar\Xi^+ + \Xi^0\bar\Xi^0} &=& \frac{1}{2} \sigma_1 \label{sig_D4}  ~,\\
   \sigma_{\bar\Delta^- n \to \Xi^-\bar\Xi^0}
   = \sigma_{\bar\Delta^0 p \to \Xi^0\bar\Xi^+} &=& \frac{1}{4} \sigma_1 ~,\\
   \sigma_{\bar\Delta^{--} p \to \Xi^-\bar\Xi^0}
   = \sigma_{\bar\Delta^+ n \to \Xi^0\bar\Xi^+} &=& \frac{3}{4} \sigma_1
\end{eqnarray}
The partial cross sections for the different outgoing channels shown with a ``+'' 
sign in (\ref{sig_nbarn_to_LbarSig0})\,-\,(\ref{sig_D4}) are equal to each other.
The expressions for the charged conjugated reaction channels are obtained by replacing all particles by the corresponding
antiparticles.

The angular differential cross sections for the $\bar N N \to \bar\Lambda \Lambda$ process
at $\sqrt{s} > 2.37$ GeV ($p_{\rm lab} > 1.830$ GeV/c) and for $\bar N N \to \bar\Lambda \Sigma, \Lambda \bar\Sigma$ processes
are chosen according to the phenomenological Regge-like fit of ref. \cite{Sadoulet:1969zh}.
At $\sqrt{s} < 2.37$ GeV, we fitted the experimental angular distributions for the 
$\bar N N \to \bar\Lambda \Lambda$ scattering of refs. \cite{Barnes:1987aw,Barnes:1989je,Barnes:1990bs,%
Barnes:1996cf,Jayet:1978yq} by the following expression: 
\begin{equation}
   \frac{d\sigma_{\bar N N \to \bar\Lambda \Lambda}}{dt}
   = [a_0 + a_1\exp(a_2(t_{\rm min}-t))]~\mu\mbox{b/GeV}^2~,    \label{dsig_dt}
\end{equation}
where $t_{\rm min}=-(p_{\bar N N}-p_{\bar\Lambda \Lambda})^2$ (in GeV$^2$),
\begin{eqnarray}
  a_0 &=& (19.1946+132202\epsilon^{2.5})/(1+42592\epsilon^{3.5})~,\\
  a_1 &=& (277.044\epsilon^{0.25}-3145.39\epsilon+4993140\epsilon^{2.8})/(1+19075.7\epsilon^3)~,\\
      && \nonumber \\
  a_2 &=&
  \cases{
     67357.8\epsilon - 257.692,   & $\epsilon \leq 0.0033$ \cr
     4.66944\ln(1.49962\epsilon), & $0.0033 < \epsilon$~.
  }
\end{eqnarray}
where $\epsilon = \sqrt{s}-2m_\Lambda$ (in GeV). For the
$\bar B B \to \bar Y Y$ processes involving $\Delta$ or $\bar\Delta$ in the initial state
and for the $\bar B B \to \bar\Xi \Xi$ processes we assume the isotropic angular differential
cross sections in the c.m. frame.

\newpage
    
\bibliography{pbarStr1}

\end{document}